\documentclass[journal=jpclcd,manuscript=letter]{achemso}

\usepackage[version=3]{mhchem} 
\usepackage{amsmath,amssymb}
\usepackage[utf8]{inputenc}
\usepackage[T1]{fontenc} 
\usepackage{mathptmx}
\usepackage{bm}
\usepackage{etoolbox}
\usepackage[labelsep=period]{caption}
\usepackage{dcolumn}
\usepackage{multirow}
\usepackage{threeparttable}
\usepackage{algorithm,algorithmic}
\usepackage{mdframed}
\usepackage{tikz}
\usetikzlibrary{shapes.geometric, arrows}
\usepackage{subfig}
\usepackage{graphicx}
\usepackage{tabularx}
\usepackage{array}
\usepackage{booktabs}
\usepackage[section]{placeins}
\usepackage{siunitx}
\usepackage{chemformula} 
\usepackage{physics} 

\newcolumntype{Y}{>{\centering\arraybackslash}X}
\newcolumntype{Z}{>{\raggedleft\arraybackslash}X}

\usepackage{caption}

\usepackage{booktabs}

\usepackage[section]{placeins}

\usepackage{siunitx}
\DeclareSIUnit\angstrom{\text {Å}}
\DeclareSIUnit\wn{\cm\tothe{-1}}

\definecolor{quotered}{rgb}{0.0,0.0,0.0}

\SectionNumbersOn

\title[NEO Multireference CI Paper 2 (Tunneling Problems)]{Computing Hydrogen Tunneling Splittings with Nuclear-Electronic Orbital Multireference Configuration Interaction}

\author{Rachel J. Stein}
\affiliation[Princeton University]{Department of Chemistry, Princeton University, Princeton, NJ 08544, United States}

\author{Christopher L. Malbon}
\affiliation[Princeton University]{Department of Chemistry, Princeton University, Princeton, NJ 08544, United States}

\author{Sharon Hammes-Schiffer}
\email{shs566@princeton.edu}
\affiliation[Princeton University]{Department of Chemistry, Princeton University, Princeton, NJ 08544, United States}

\begin{document}

\begin{tocentry}
\centering    
\includegraphics[width=3.25in]{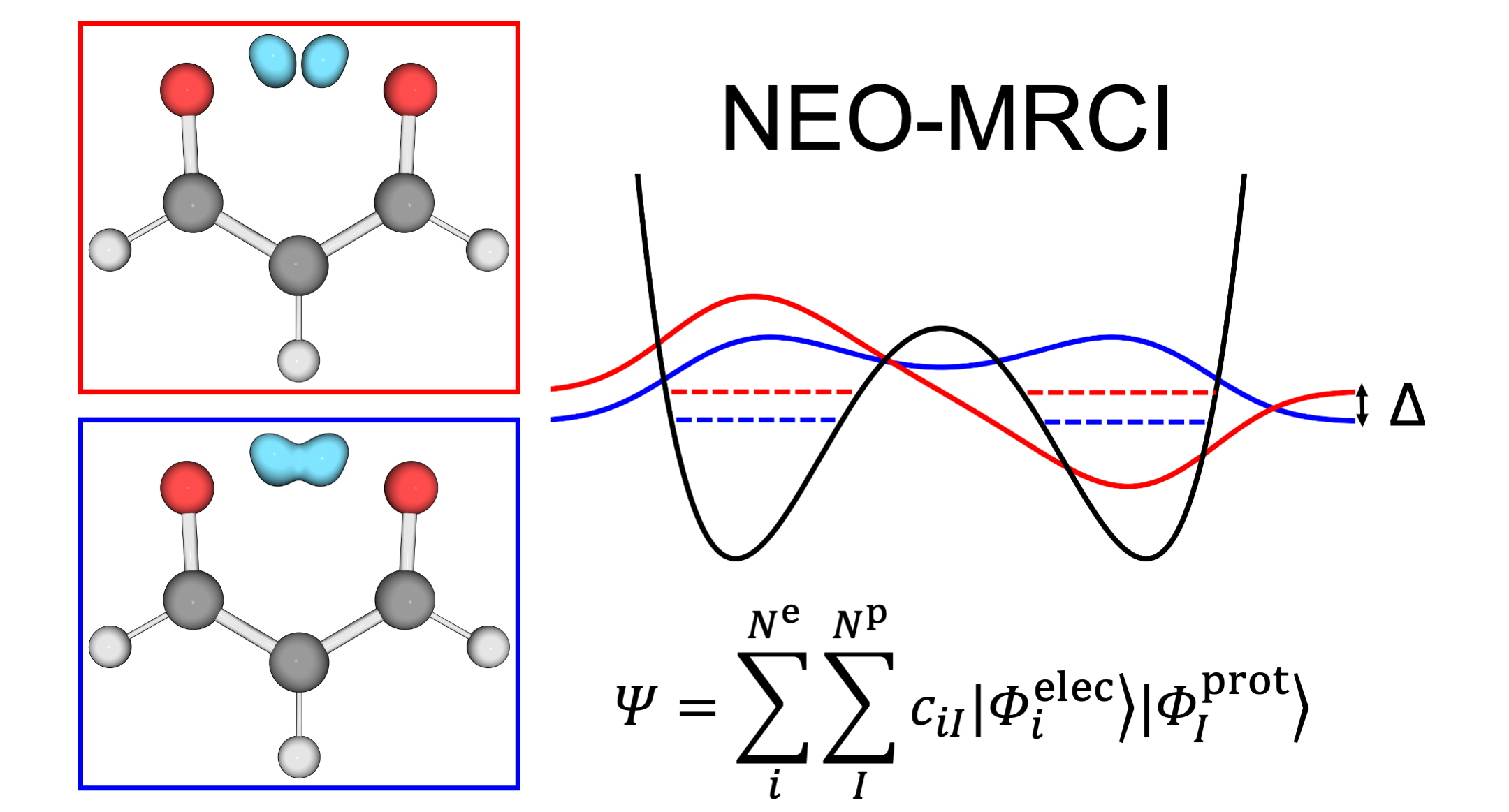}

\end{tocentry}

\begin{abstract}

Hydrogen tunneling is an important process that impacts reaction rates and molecular spectra.
Describing and understanding this process requires a quantum mechanical treatment of the transferring hydrogen.
The nuclear-electronic orbital (NEO) approach treats specified nuclei quantum mechanically on the same level as electrons and has recently been implemented at the multireference configuration interaction (MRCI) wavefunction level.
The NEO-MRCI method includes both the static correlation necessary to describe hydrogen tunneling and the electron-proton dynamic correlation required for computing quantitatively accurate nuclear-electronic vibronic states.
Herein, the NEO-MRCI method is used to compute the nuclear-electronic wavefunctions and corresponding vibronic energies for four hydrogen tunneling systems at fixed geometries for a range of donor-acceptor distances.
Comparison of the NEO-MRCI results to numerically exact grid-based calculations  shows that the NEO-MRCI method can be used to obtain accurate hydrogen and deuterium tunneling splittings {\color{quotered}at fixed geometries}.
Thus, this work presents an {\color{quotered}important component for }studying hydrogen tunneling systems.
\end{abstract}

\begin{figure}
    \centering
    \includegraphics[width=3.25in]{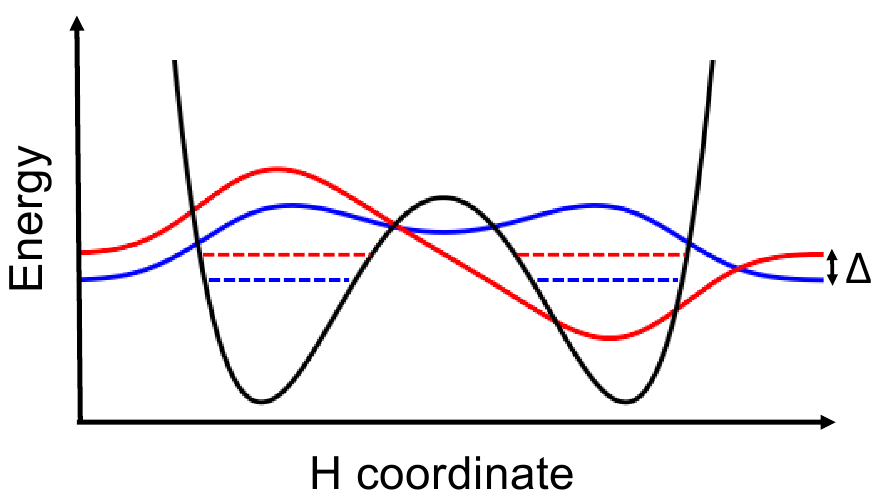}
    \caption{Schematic one-dimensional double-well proton potential energy profile and associated symmetric (blue) and antisymmetric (red) proton vibrational wavefunctions with vibrational state energies indicated by dotted lines. The tunneling splitting, $\Delta$, is the difference in energy between these two vibrational states. This one-dimensional picture is for visualization purposes; all tunneling splitting calculations in this work were calculated in three dimensions.}
    \label{fig:tunsplit_schematic}
\end{figure}

Hydrogen tunneling and proton transfer play important roles in a wide range of chemical processes in molecular, biological, and electrochemical systems.
A prime example is the prevalence of hydrogen tunneling in proton-coupled electron transfer (PCET) in enzymatic and electrocatalytic\cite{ChaMurrayKlinman1989,HogansonBabcock1997,KnappRickertKlinman2002,StubbeNoceraYeeChang2003,MagnusonAnderlundEtAl2009,SHSStuchebrukhov2010,WarrenTronicMayer2010,DuboisBullock2011}
reactions. Hydrogen tunneling splittings can be measured spectroscopically,\cite{BaughcumDuerst1984,FirthLeopold1991,BabaTanaka1999,BirerHavenith2009,VdovinSlenczka2009,TalbotJohnson2020} providing information about the structural and dynamical properties.
Additionally, hydrogen isotope effects on tunneling splittings can offer valuable insights into reaction mechanisms.\cite{SenKohen2010,HamaWatanabe2015,KlinmanOffenbacher2018,SHS2025}
The ground state tunneling splitting is typically defined as the energy difference between the lowest two vibronic states corresponding to symmetric and antisymmetric bilobal wavefunctions (Fig. \ref{fig:tunsplit_schematic}). 
For electronically adiabatic systems, these two vibronic states correspond to the lowest two proton vibrational states on the electronic ground state, where the proton is moving on a symmetric (or nearly symmetric) double-well potential energy surface. 
Given the widespread importance of hydrogen tunneling and the sensitivity of tunneling to the underlying potential energy surface, the development of high-level theoretical methods to compute precise proton vibrational wavefunctions, accurate tunneling splittings, and isotope effects is desirable.

The calculation of tunneling splittings is notoriously challenging. 
These splittings depend strongly on the electronic potential energy surface, especially the width and height of the tunneling barrier.
An accurate description of hydrogen tunneling requires the incorporation of the nuclear quantum effects of the tunneling proton and its coupling to the electronic wavefunction. 
A wide range of theoretical methods have been developed to model tunneling processes, such as path integral methods,\cite{Mil'nikoovNakamura2001,RichardsonAlthorpe2011,TreninsMeuserBertschiVavourakisFlutschRichardson2023} diffusion Monte Carlo\cite{GregoryClary1995,FortenberryFrancisco2015, DiRisioFinneyMcCoy2022, FinneyMcCoy2024}, and the multiconfiguration time-dependent Hartree approach.\cite{Coutinho-NetoVielManthe2004,SchröderGattiMeyer2011,SuzukiKono2023}
The nuclear-electronic orbital (NEO) approach\cite{WebbIordanovSHS2002,SHS2021} treats specified nuclei, typically protons, as well as all electrons quantum mechanically, thereby offering an alternative strategy.
Many analogues of conventional electronic structure methods have been implemented within the NEO framework, including density functional theory,\cite{PakChakrabortySHS2007,YangSHS2017,BrorsenYangSHS2017} time-dependent density functional theory (TDDFT),\cite{YangCulpittSHS2018,CulpittSHS2019} coupled-cluster theory,\cite{PavoševićCulpittSHS2019,PavoševićTaoSHS2021,PavoševićSHS2022} density matrix renormalization group,\cite{MuoloBaiardiFeldmannReiher2020,FeldmannMuoloBaiardiReiher2022} and the complete active space self-consistent field method.\cite{WebbIordanovSHS2002,FajenBrorsen2021}

To capture the bilobal nature of the hydrogen nuclear wavefunction in tunneling systems within the NEO framework, a multireference method is required.\cite{PakSHS2004,PakSwalinaWebbSHS2004,SwalinaPakSHS2005,SkonePakSHS2005,YuSHS2020,DickinsonYuSHS2023,GarnerUpadhyayLiSHS2024}
Typically, the transferring hydrogen nucleus is represented by two basis function centers, each with a protonic and an electronic basis set.\cite{PakSHS2004,PakSwalinaWebbSHS2004,SwalinaPakSHS2005,SkonePakSHS2005,YuSHS2020,DickinsonYuSHS2023,GarnerUpadhyayLiSHS2024} The NEO nonorthogonal configuration interaction (NEO-NOCI) method has been used to obtain tunneling splittings of simple systems,\cite{SkonePakSHS2005} but the formalism would need to be extended significantly to be more generally applicable.
The NEO multistate density functional theory (NEO-MSDFT) method has been used to study larger tunneling systems, in some cases with multiple tunneling protons.\cite{YuSHS2020,DickinsonYuSHS2023} 
The NEO-MSDFT method depends on tunable parameters and approximate functionals to obtain quantitatively accurate tunneling splittings, but it is useful for real-time nuclear-electronic quantum dynamics simulations in large chemical and biological systems.\cite{YuRoySHS2022,DickinsonSHS2024}
A more rigorous wavefunction-based method that is systematically improvable and can provide accurate tunneling splittings for molecular systems without free parameters is desirable.

In this Letter, we apply the recently introduced NEO-MRCI method\cite{MalbonSHS2025} to several hydrogen tunneling systems, demonstrating it can be used to calculate accurate {\color{quotered}hydrogen }tunneling splittings for fixed geometries. {\color{quotered}This demonstration is a key step toward computing tunneling splittings that include the other nuclear motions to enable  comparison to experimental measurements.}
Analogous to the conventional MRCI method,\cite{Lishchka2018,Matsika2021} the NEO-MRCI method is parameter-free and offers a natural description of excited states, which are vibronic states within the NEO framework. As a multireference method, it is well-suited to tunneling problems within the NEO framework.
In contrast to most other methods for computing tunneling splittings, the electronic and nuclear wavefunctions are computed simultaneously with no Born-Oppenheimer separation between the electrons and the transferring proton.
The four hydrogen tunneling systems studied in this work are \ce{HeHHe+}, \ce{OCHCO+}, \ce{FHF-}, and malonaldehyde.
The \ce{HeHHe+}, \ce{OCHCO+}, and \ce{FHF-} systems exhibit hydrogen tunneling at extended donor-acceptor distances.\cite{TerrillNesbitt2010}
Malonaldehyde is a prototypical system for studying intramolecular hydrogen tunneling and has been studied widely both experimentally and theoretically.\cite{BaughcumDuerst1984, ShidaBarbaraAlmlöf1989, FirthLeopold1991, BabaTanaka1999,Coutinho-NetoVielManthe2004,SchröderGattiMeyer2011,MizukamiHabershonTew2014,VaillantWalesAlthorpe2018,LawrenceDusekRichardson2023,LauvergnatNauts2023,BaumannTreninsRichardson2025}

To benchmark the NEO-MRCI results, the three-dimensional Fourier Grid Hamiltonian \newline (FGH)\cite{MarstonBalintKurti1989,WebbSHS2000} method was used to compute numerically exact reference tunneling splittings {\color{quotered}at fixed geometries.} 
The FGH method solves for the proton vibrational wavefunctions given a three-dimensional electronic potential energy surface generated by moving the transferring hydrogen nucleus on a grid while the other nuclei remain fixed.
In this work, the energy at each grid point was computed with the conventional electronic coupled-cluster singles and doubles (CCSD) method. The aug-cc-pVTZ\cite{KendallDunningHarrison1992} electronic basis set was used for \ce{HeHHe+}, \ce{OCHCO+}, and \ce{FHF-}, and the cc-pVTZ\cite{Dunning1989} electronic basis set was used for malonaldehyde. 
The resulting proton vibrational wavefunctions and tunneling splittings are directly comparable to NEO calculations for fixed geometries of the classical nuclei. 
{\color{quotered} We chose to use CCSD for the FGH benchmark because it produces an accurate electronic ground-state potential energy surface. In general, it is challenging to achieve the same electronic structure for conventional and NEO calculations, and such differences in electronic structure could introduce minor discrepancies between the NEO and FGH results.
Note that the tunneling splittings computed for fixed geometries are not comparable to experimentally measured tunneling splittings because the hydrogen tunneling mode is typically coupled to the vibrational modes of other nuclei.}
The effect of this coupling can be included using a method such as vibronic coupling theory.\cite{HazraSkoneSHS2009}
To benchmark the NEO-MRCI method for electronically adiabatic hydrogen tunneling splittings, however, the FGH results provide a numerically exact reference.

The NEO-MRCI wavefunction is an expansion of NEO configurations:
\begin{equation}
    \Psi^\text{NEO} (\mathbf{R}^\text{c}) = \sum_\mu c_\mu (\mathbf{R}^\text{c}) \psi_\mu^\text{NEO} (\mathbf{R}^\text{c})    
\end{equation}
where each configuration $\psi^\text{NEO}_\mu$ is the product of electronic and protonic Slater determinants
\begin{equation}
    \psi^\text{NEO}_\mu (\mathbf{R}^\text{c}) = \lvert \Phi_{i(\mu)}^\text{e} (\mathbf{R}^\text{c})\rangle \lvert \Phi_{I(\mu)}^\text{p} (\mathbf{R}^\text{c})\rangle.
\end{equation}
Here ${\bf R}^\text{c}$ are the coordinates of the classically treated nuclei and are fixed during the NEO-MRCI calculation.
$\lvert \Phi_{i(\mu)}^\text{e}\rangle$ and $\lvert \Phi_{I(\mu)}^\text{p} \rangle$ are the $i$-th electronic determinant and $I$-th protonic determinant, respectively, that form the $\mu$-th NEO configuration, $\psi^\text{NEO}_\mu$.
The electronic and protonic determinants are composed of electronic and protonic orbitals optimized in the NEO-MCSCF reference space defined by electronic and protonic active spaces.\cite{MalbonSHS2025}
In this work, the orbitals are optimized for the lowest two vibronic states using the state averaged NEO-MCSCF (NEO-SA-MCSCF) method.
The NEO-MR-SD$_\text{en}$CI expansion is constructed from single electron, single proton, and double electron-proton excitations from this reference space.\cite{MalbonSHS2025}
{\color{quotered}Here, S refers to all single excitations (i.e., single electron and single proton), and D$_\text{en}$ refers to double electron-proton excitations.
To compute hydrogen tunneling splittings, which are predominantly vibrational excitations, double electronic excitations are not expected to be critical and would increase the computational cost significantly.}

We calculated hydrogen tunneling splittings for \ce{HeHHe+}, \ce{FHF-}, and \ce{OCHCO+} at various He---He, F---F, and C---C distances and at two geometries of malonaldehyde.
The geometry used for \ce{OCHCO+} was obtained from earlier work\cite{YuSHS2020}, and the C---O distances for each CO moiety were kept fixed as the C---C distance was increased.
For malonaldehyde, an average of reactant and product structures was used, where the reactant and product correspond to the hydrogen optimized near the donor or acceptor, respectively.
We examined two malonaldehyde geometries, where one structure was generated using the equilibrium reactant and product geometries, corresponding to an O---O distance of 2.617 Å, rounded to 2.62 Å in the rest of the paper, and the other
structure was generated with the O---O distance constrained to 2.55 Å during optimization of the reactant and product structures. These geometry optimizations were performed at the conventional CCSD/cc-pVTZ level of theory, and the coordinates are given in in Section 8.4 of the SI.

In all NEO calculations, the hydrogen basis function center positions were optimized with conventional CCSD.
For each geometry, the proton was optimized near the donor atom with all other nuclei held fixed.
One proton basis function center was placed at the optimized position, and the other was placed at the mirror image near the acceptor for symmetric systems.
For non-symmetric systems, the position of the second proton basis function center would be determined by a second optimization near the acceptor.
This procedure results in proton basis function centers positioned at the minima of the electronically adiabatic CCSD double-well potential energy surface.
As shown from the examples below, optimizing the basis function center positions at the conventional CCSD level provides a reliable method for computing hydrogen tunneling splittings with the NEO-MRCI method.

Accurate tunneling splittings require large electronic and protonic basis sets,
in accordance with previous work demonstrating that accurate vibronic excitation energies computed with NEO-TDDFT require large electronic basis sets.\cite{CulpittSHS2019}
The \ce{HeHHe+} and \ce{FHF-} NEO calculations used the even-tempered 8s8p8d8f\cite{YangSHS2017} protonic basis set.
To reduce computational cost, the \ce{OCHCO+} and malonaldehyde NEO calculations used an even-tempered 5s5p5d5f protonic basis set, where $\alpha$ = 2.8 and $\beta$ = 1.8459 according to the standard formula,\cite{Helgaker2000} covering the same exponent range as 8s8p8d8f.
We show that the same tunneling splitting is obtained for \ce{HeHHe+} using the 8s8p8d8f and 5s5p5d5f basis sets (Table S5).
All NEO calculations used the cc-pV5Z* electronic basis set for the quantum proton. 
The asterisk indicates that the most diffuse basis function from the core contraction is used to represent the 1s orbital, while the rest of the primitives from the core contraction are removed.
Previous work has shown that NEO calculations should not use the same electronic basis sets as conventional calculations because the quantized proton is no longer a point charge.\cite{MalbonSHS2025}
Results with different protonic and electronic basis sets are provided in the SI to show convergence (Tables S1-S7).
The remaining atoms in the NEO calculations were represented with the same electronic basis set used in the corresponding FGH calculation: aug-cc-pVTZ for \ce{HeHHe+}, \ce{FHF-}, and \ce{OCHCO+},  and cc-pVTZ for malonaldehyde.
These same electronic basis sets were used for all atoms in the conventional CCSD proton basis function center optimizations.

The molecular orbitals were optimized in a NEO-SA-MCSCF calculation, weighing the two lowest vibronic states equally.
To ensure the electronic and protonic molecular orbitals are symmetric, the initial guess orbitals were generated from a two-proton NEO-HF calculation.
Although this molecule containing an additional proton is unphysical, the resulting orbitals serve as a reasonable initial guess for the NEO-SA-MCSCF calculation.
Full protonic active spaces were used for all systems in this work. {\color{quotered} A smaller protonic active space would most likely be sufficient, but the computational cost of using a full protonic active space is negligible.} 
A minimal (2e, 2o) active space, containing orbitals of in-phase $\sigma$ and $\sigma$* character, was used for \ce{FHF-}, \ce{OCHCO+}, and malonaldehyde (Figures S2-S4). {\color{quotered}These orbitals correspond to bonding and antibonding of the hydrogen with the donor and acceptor along the molecular axis. To compute hydrogen tunneling splittings, it is critical to maintain a balanced treatment and symmetry across the tunneling barrier.}
Interestingly, the same (2e, 2o) active space was able to predict accurate tunneling splittings in three different systems.
\ce{HeHHe+} is different from the other systems in that it resembles a bare proton transferring between the donor and acceptor with negligible electron density in the central region of the molecule (Fig. S1). 
Because the chemical bonding structure is qualitatively different for \ce{HeHHe+} in that the covalent sigma bond between helium and hydrogen is not formed, a single-reference electronic treatment within NEO-CASSCF is sufficient for this system.
The NEO calculations to obtain deuterium tunneling splittings for \ce{FDF-} and malonaldehyde were performed with the same active spaces and basis sets as the hydrogen calculations. {\color{quotered}Because we are using large, even-tempered protonic basis sets, scaling the exponents to account for the greater mass of deuterium has a negligible effect (Table S6).}


\begin{figure}
    \centering
    \includegraphics[width=3.25in]{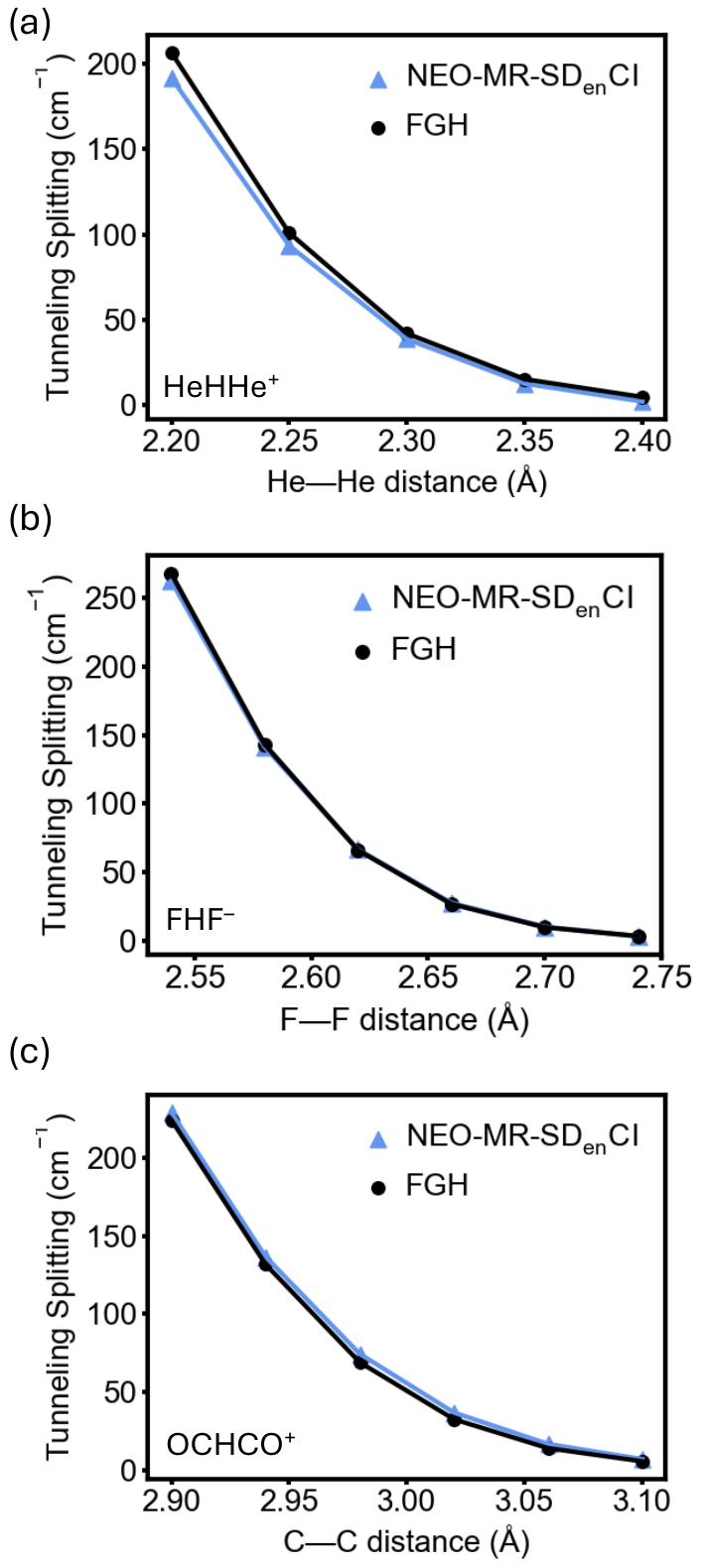}
    \caption{NEO-MR-SD$_\text{en}$CI hydrogen tunneling splittings compared with the reference FGH results for (a) \ce{HeHHe+}, (b) \ce{FHF-}, and (c) \ce{OCHCO+} at different donor-acceptor distances.
    The reference FGH results were computed at the conventional CCSD/aug-cc-pVTZ level of theory. The numerical values are provided in Tables S1-S3.}
    \label{fig:tunsplit_plot}
\end{figure}

\begin{figure}
    \centering
    \includegraphics[width=6.5in]{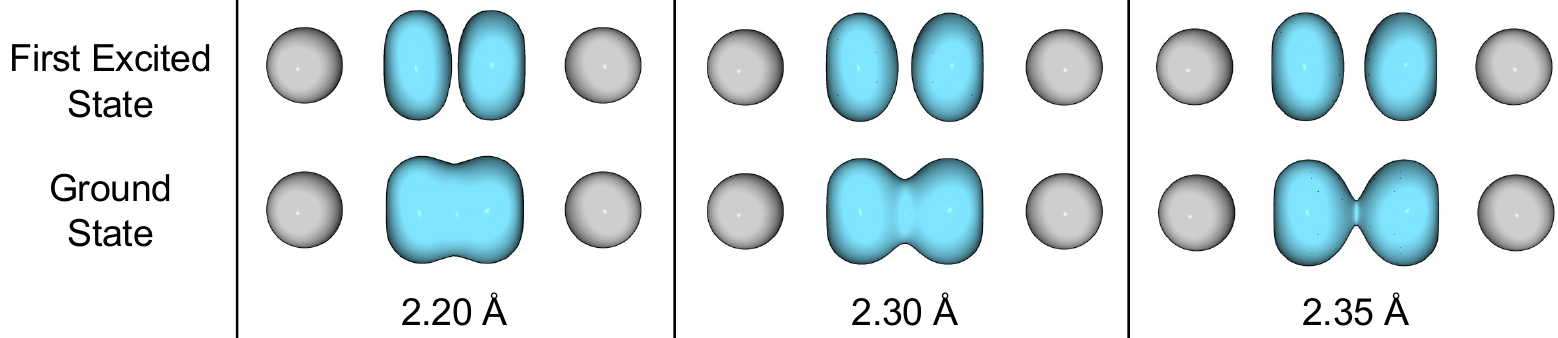}
    \caption{NEO-MR-SD$_\text{en}$CI protonic densities for \ce{HeHHe+} at the indicated He---He distances. The He atoms are shown in light gray, and the protonic densities are shown in cyan  with an isovalue of 0.04 e $\text{Bohr}^{-3}$. The analogous protonic densities for the other systems are shown in Figures S5 and S6.}
    \label{fig:hehhe_densities}
\end{figure}

The hydrogen tunneling splittings calculated for \ce{HeHHe+} at the NEO-MR-SD$_\text{en}$CI level agree quantitatively with the tunneling splittings calculated using the numerically exact FGH method {\color{quotered} at fixed He---He distances.} 
As expected, the tunneling splitting decreases with increased He---He distance due to the increasing barrier height and width (Fig.~\ref{fig:tunsplit_plot}a). 
The NEO-MR-SD$_\text{en}$CI protonic densities also exhibit the correct qualitative behavior (Fig.~\ref{fig:hehhe_densities}).
The ground and first excited state protonic densities are bilobal, and 
the ground state proton vibrational density has no nodes, whereas the first excited state proton density contains a single node. 
The other two linear systems behave similarly to \ce{HeHHe+} despite having more complex electronic structures.
The NEO-MR-SD$_\text{en}$CI tunneling splittings of \ce{FHF-} and \ce{OCHCO+} decrease as the donor-acceptor distance increases and quantitatively agree with the FGH tunneling splittings (Fig.~\ref{fig:tunsplit_plot}b and c).
{\color{quotered}Tables S8 and S9 provide information about the computational cost for representative calculations.}

\begin{figure}
    \centering
    
    \includegraphics[width=6.5in]{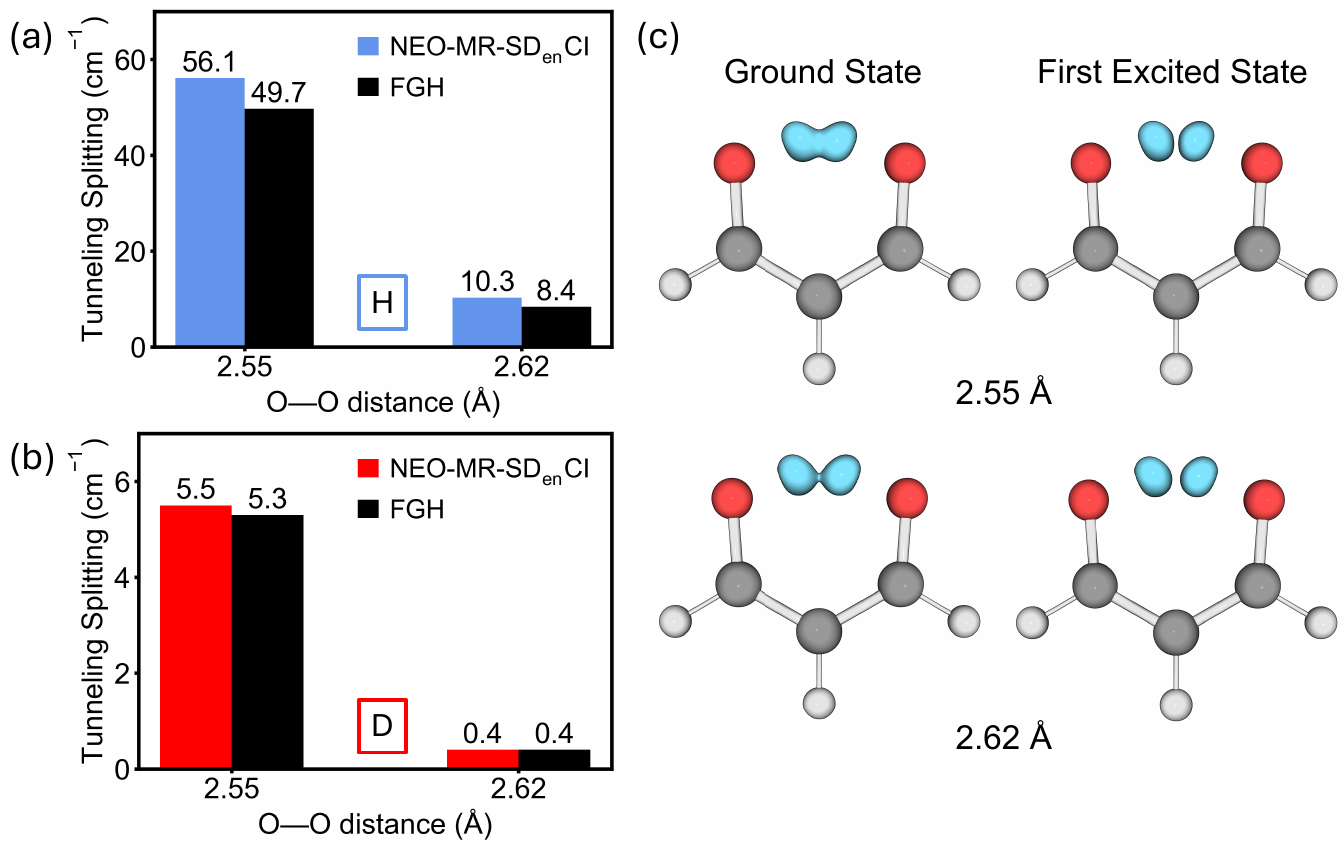}
    \caption{NEO-MR-SD$_\text{en}$CI tunneling splittings for malonaldehyde at two different donor-acceptor distances compared to the FGH results with (a) hydrogen and (b) deuterium. 
    The FGH results were computed at the CCSD/cc-pVTZ level of theory. (c) Protonic densities for hydrogen at the indicated O---O distances. The protonic densities are shown in cyan with an isovalue of 0.04 e $\text{Bohr}^{-3}$,  and the other atoms are shown in dark gray (carbon), red (oxygen), and light gray (hydrogen).}
    \label{fig:malon}
\end{figure}

The malonaldehyde results further demonstrate the broad applicability of the NEO-MRCI method to tunneling problems.
The tunneling splittings for the two malonaldehyde geometries corresponding to different O---O distances are shown in Figure \ref{fig:malon}a.
NEO-MR-SD$_\text{en}$CI tunneling splittings for both of these geometries are in excellent agreement with the FGH benchmarks.
The protonic densities are bilobal with the appropriate nodal structure but are slightly squeezed to form a "V" shape (Fig.~\ref{fig:malon}c).
This behavior is consistent with previous studies demonstrating that malonaldehyde follows a non-linear tunneling path.\cite{ShidaBarbaraAlmlöf1989,TautermannLiedl2002}
This level of agreement in the tunneling splittings for a larger, non-linear molecule supports the application of NEO-MR-SD$_\text{en}$CI for studying other interesting hydrogen tunneling systems.
{\color{quotered}However, we emphasize that the experimental tunneling splitting for malonaldehyde is 21.6 cm$^{-1}$,\cite{BaughcumDuerst1984,FirthLeopold1991,BabaTanaka1999} which has been computed accurately with other theoretical methods that include all nuclear motions.\cite{MizukamiHabershonTew2014,LawrenceDusekRichardson2023,LauvergnatNauts2023,BaumannTreninsRichardson2025} Within the NEO framework, NEO-MR-SD$_\text{en}$CI can be combined with an approach such as vibronic coupling theory to incorporate the other nuclear motions and their coupling to the hydrogen tunneling mode.\cite{HazraSkoneSHS2009} This type of combined approach is a direction for future research.}

\begin{figure}
    \centering
    
    \includegraphics[width=3.25in]{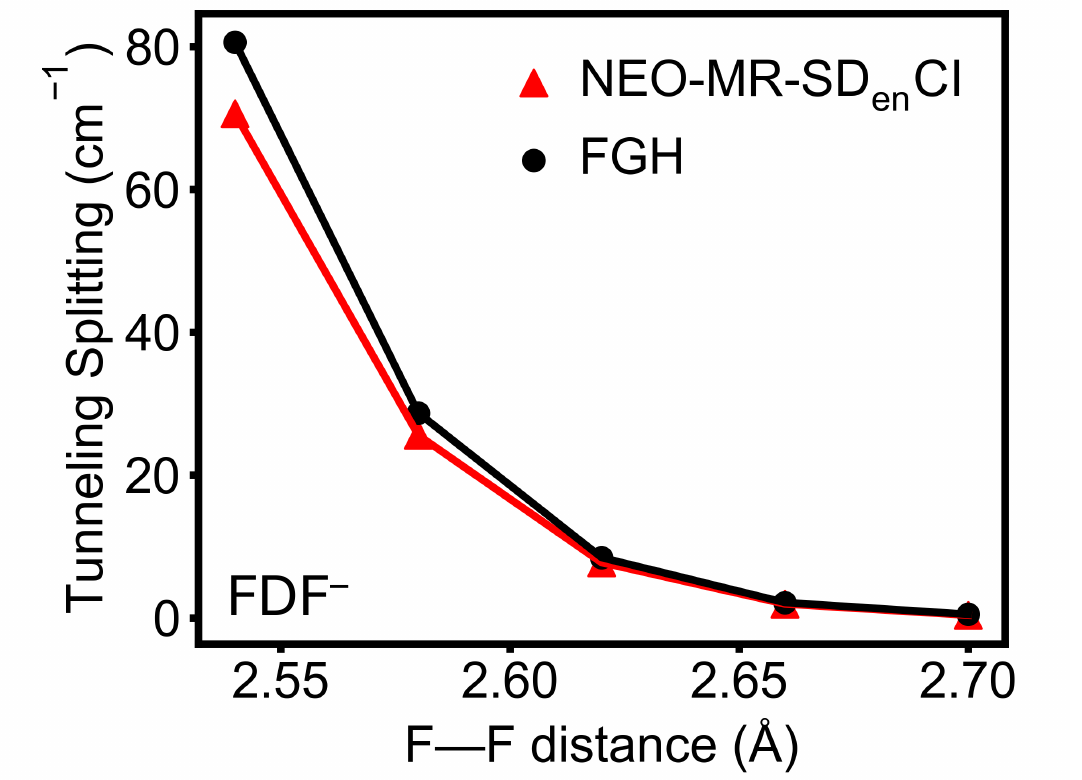}
    \caption{NEO-MR-SD$_\text{en}$CI deuterium tunneling splittings compared with the reference FGH results for \ce{FDF-} at different donor-acceptor distances.
    The reference FGH results were computed at the conventional CCSD/aug-cc-pVTZ level of theory. The numerical values are provided in Table S2.}
    \label{fig:deut_FHF}
\end{figure}

The NEO-MR-SD$_\text{en}$CI method can also be used to calculate deuterium tunneling splittings.
We calculated the deuterium tunneling splittings for the \ce{FDF-} and malonaldehyde systems.
As shown in Figure \ref{fig:deut_FHF}, the NEO-MR-SD$_\text{en}$CI deterium tunneling splittings are in excellent agreement with the FGH results at a range of donor-acceptor distances for \ce{FDF-}.
The deuterium tunneling splittings are much smaller than the hydrogen tunneling splittings, as expected for a heavier quantum nucleus.
The NEO-MR-SD$_\text{en}$CI deuterium tunneling splittings for malonaldehyde also agree well with the FGH reference results and follow the same trend as the \ce{FDF-} deuterium tunneling splittings (Fig. \ref{fig:malon}b).

Hydrogen tunneling systems are a challenge to model due to their multireference nature in the NEO framework and the small magnitudes of typical tunneling splittings.
In this Letter, we have demonstrated that NEO-MRCI is an effective {\color{quotered} component for the calculation of} tunneling splittings for molecular systems. 
NEO-MRCI does not require any parameters, accounts for the multireference nature of the system, and captures the correlation energy required to obtain quantitatively accurate results.
Moreover, the NEO-MRCI method can describe hydrogen tunneling in systems with both linear and non-linear tunneling paths, such as malonaldehyde.
We have devised an efficient and unbiased strategy using conventional CCSD optimizations to determine the proton basis function center positions.
We demonstrate that relatively large electronic and protonic basis sets, as well as a (2e, 2o) active space consisting of in-phase $\sigma$ and $\sigma$* orbitals, can effectively describe the vibronic tunneling states in a variety of systems.
With these insights, the NEO-MRCI approach produces accurate hydrogen tunneling splittings in applications to four different molecular systems at {\color{quotered} fixed geometries with} a range of donor-acceptor distances.
Additionally, this scheme produces accurate deuterium tunneling splittings.
These results show that the NEO-MRCI approach is {\color{quotered}an important step toward} investigating a wide range of hydrogen tunneling systems in chemistry and biology {\color{quotered}with multicomponent quantum chemistry.}

\begin{acknowledgement}
The authors thank Dr. Scott Garner, Dr. Jonathan Fetherolf, Dr. Eno Paenurk, Dr. Chiara Aieta, Joseph Dickinson, Mathew Chow, Millan Welman, and Rowan Goudy for useful discussions. This work was supported by the National Science Foundation
Grant No. CHE-2408934.

\end{acknowledgement}

\begin{suppinfo}
The Supporting Information is available free of charge. Tunneling splitting data, basis set convergence data, electron density, electronic active spaces, \ce{OCHCO+} and \ce{FHF-} protonic densities, computational cost analysis, molecular coordinates.

\end{suppinfo}


\bibliography{neo_multirefci}

\providecommand{\latin}[1]{#1}
\makeatletter
\providecommand{\doi}
  {\begingroup\let\do\@makeother\dospecials
  \catcode`\{=1 \catcode`\}=2 \doi@aux}
\providecommand{\doi@aux}[1]{\endgroup\texttt{#1}}
\makeatother
\providecommand*\mcitethebibliography{\thebibliography}
\csname @ifundefined\endcsname{endmcitethebibliography}  {\let\endmcitethebibliography\endthebibliography}{}
\begin{mcitethebibliography}{68}
\providecommand*\natexlab[1]{#1}
\providecommand*\mciteSetBstSublistMode[1]{}
\providecommand*\mciteSetBstMaxWidthForm[2]{}
\providecommand*\mciteBstWouldAddEndPuncttrue
  {\def\EndOfBibitem{\unskip.}}
\providecommand*\mciteBstWouldAddEndPunctfalse
  {\let\EndOfBibitem\relax}
\providecommand*\mciteSetBstMidEndSepPunct[3]{}
\providecommand*\mciteSetBstSublistLabelBeginEnd[3]{}
\providecommand*\EndOfBibitem{}
\mciteSetBstSublistMode{f}
\mciteSetBstMaxWidthForm{subitem}{(\alph{mcitesubitemcount})}
\mciteSetBstSublistLabelBeginEnd
  {\mcitemaxwidthsubitemform\space}
  {\relax}
  {\relax}

\bibitem[Cha \latin{et~al.}(1989)Cha, Murray, and Klinman]{ChaMurrayKlinman1989}
Cha,~Y.; Murray,~C.~J.; Klinman,~J.~P. Hydrogen Tunneling in Enzyme Reactions. \emph{Science} \textbf{1989}, \emph{243}, 1325--1330\relax
\mciteBstWouldAddEndPuncttrue
\mciteSetBstMidEndSepPunct{\mcitedefaultmidpunct}
{\mcitedefaultendpunct}{\mcitedefaultseppunct}\relax
\EndOfBibitem
\bibitem[Hoganson and Babcock(1997)Hoganson, and Babcock]{HogansonBabcock1997}
Hoganson,~C.~W.; Babcock,~G.~T. A Metalloradical Mechanism for the Generation of Oxygen from Water in Photosynthesis. \emph{Science} \textbf{1997}, \emph{277}, 1953--1956\relax
\mciteBstWouldAddEndPuncttrue
\mciteSetBstMidEndSepPunct{\mcitedefaultmidpunct}
{\mcitedefaultendpunct}{\mcitedefaultseppunct}\relax
\EndOfBibitem
\bibitem[Knapp \latin{et~al.}(2002)Knapp, Rickert, and Klinman]{KnappRickertKlinman2002}
Knapp,~M.~J.; Rickert,~K.; Klinman,~J.~P. Temperature-Dependent Isotope Effects in Soybean Lipoxygenase-1: Correlating Hydrogen Tunneling with Protein Dynamics. \emph{J. Am. Chem. Soc.} \textbf{2002}, \emph{124}, 3865--3874\relax
\mciteBstWouldAddEndPuncttrue
\mciteSetBstMidEndSepPunct{\mcitedefaultmidpunct}
{\mcitedefaultendpunct}{\mcitedefaultseppunct}\relax
\EndOfBibitem
\bibitem[Stubbe \latin{et~al.}(2003)Stubbe, Nocera, Yee, and Chang]{StubbeNoceraYeeChang2003}
Stubbe,~J.; Nocera,~D.~G.; Yee,~C.~S.; Chang,~M. C.~Y. Radical Initiation in the Class I Ribonucleotide Reductase: Long-Range Proton-Coupled Electron Transfer? \emph{Chem. Rev.} \textbf{2003}, \emph{103}, 2167--2202\relax
\mciteBstWouldAddEndPuncttrue
\mciteSetBstMidEndSepPunct{\mcitedefaultmidpunct}
{\mcitedefaultendpunct}{\mcitedefaultseppunct}\relax
\EndOfBibitem
\bibitem[Magnuson \latin{et~al.}(2009)Magnuson, Anderlund, Johansson, Lindblad, Lomoth, Polivka, Ott, Stensjö, Styring, Sundström, and Hammarström]{MagnusonAnderlundEtAl2009}
Magnuson,~A.; Anderlund,~M.; Johansson,~O.; Lindblad,~P.; Lomoth,~R.; Polivka,~T.; Ott,~S.; Stensjö,~K.; Styring,~S.; Sundström,~V. \latin{et~al.}  Biomimetic and Microbial Approaches to Solar Fuel Generation. \emph{Acc. Chem. Res.} \textbf{2009}, \emph{42}, 1899--1909\relax
\mciteBstWouldAddEndPuncttrue
\mciteSetBstMidEndSepPunct{\mcitedefaultmidpunct}
{\mcitedefaultendpunct}{\mcitedefaultseppunct}\relax
\EndOfBibitem
\bibitem[Hammes-Schiffer and Stuchebrukhov(2010)Hammes-Schiffer, and Stuchebrukhov]{SHSStuchebrukhov2010}
Hammes-Schiffer,~S.; Stuchebrukhov,~A.~A. Theory of Coupled Electron and Proton Transfer Reactions. \emph{Chem. Rev.} \textbf{2010}, \emph{110}, 6939--6960\relax
\mciteBstWouldAddEndPuncttrue
\mciteSetBstMidEndSepPunct{\mcitedefaultmidpunct}
{\mcitedefaultendpunct}{\mcitedefaultseppunct}\relax
\EndOfBibitem
\bibitem[Warren \latin{et~al.}(2010)Warren, Tronic, and Mayer]{WarrenTronicMayer2010}
Warren,~J.~J.; Tronic,~T.~A.; Mayer,~J.~M. Thermochemistry of Proton-Coupled Electron Transfer Reagents and its Implications. \emph{Chem. Rev.} \textbf{2010}, \emph{110}, 6961--7001\relax
\mciteBstWouldAddEndPuncttrue
\mciteSetBstMidEndSepPunct{\mcitedefaultmidpunct}
{\mcitedefaultendpunct}{\mcitedefaultseppunct}\relax
\EndOfBibitem
\bibitem[DuBois and Bullock(2011)DuBois, and Bullock]{DuboisBullock2011}
DuBois,~D.~L.; Bullock,~R.~M. Molecular Electrocatalysts for the Oxidation of Hydrogen and the Production of Hydrogen – The Role of Pendant Amines as Proton Relays. \emph{Eur. J. Inorg. Chem.} \textbf{2011}, \emph{2011}, 1017--1027\relax
\mciteBstWouldAddEndPuncttrue
\mciteSetBstMidEndSepPunct{\mcitedefaultmidpunct}
{\mcitedefaultendpunct}{\mcitedefaultseppunct}\relax
\EndOfBibitem
\bibitem[Baughcum \latin{et~al.}(1984)Baughcum, Smith, Wilson, and Duerst]{BaughcumDuerst1984}
Baughcum,~S.~L.; Smith,~Z.; Wilson,~E.~B.; Duerst,~R.~W. Microwave spectroscopic study of malonaldehyde. 3. Vibration-rotation interaction and one-dimensional model for proton tunneling. \emph{J. Am. Chem. Soc.} \textbf{1984}, \emph{106}, 2260--2265\relax
\mciteBstWouldAddEndPuncttrue
\mciteSetBstMidEndSepPunct{\mcitedefaultmidpunct}
{\mcitedefaultendpunct}{\mcitedefaultseppunct}\relax
\EndOfBibitem
\bibitem[Firth \latin{et~al.}(1991)Firth, Beyer, Dvorak, Reeve, Grushow, and Leopold]{FirthLeopold1991}
Firth,~D.~W.; Beyer,~K.; Dvorak,~M.~A.; Reeve,~S.~W.; Grushow,~A.; Leopold,~K.~R. Tunable far‐infrared spectroscopy of malonaldehyde. \emph{J. Chem. Phys.} \textbf{1991}, \emph{94}, 1812--1819\relax
\mciteBstWouldAddEndPuncttrue
\mciteSetBstMidEndSepPunct{\mcitedefaultmidpunct}
{\mcitedefaultendpunct}{\mcitedefaultseppunct}\relax
\EndOfBibitem
\bibitem[Baba \latin{et~al.}(1999)Baba, Tanaka, Morino, Yamada, and Tanaka]{BabaTanaka1999}
Baba,~T.; Tanaka,~T.; Morino,~I.; Yamada,~K. M.~T.; Tanaka,~K. Detection of the tunneling-rotation transitions of malonaldehyde in the submillimeter-wave region. \emph{J. Chem. Phys.} \textbf{1999}, \emph{110}, 4131--4133\relax
\mciteBstWouldAddEndPuncttrue
\mciteSetBstMidEndSepPunct{\mcitedefaultmidpunct}
{\mcitedefaultendpunct}{\mcitedefaultseppunct}\relax
\EndOfBibitem
\bibitem[Birer and Havenith(2009)Birer, and Havenith]{BirerHavenith2009}
Birer,~{\"O}.; Havenith,~M. High-Resolution Infrared Spectroscopy of the Formic Acid Dimer. \emph{Annu. Rev. Phys. Chem.} \textbf{2009}, \emph{60}, 263--275\relax
\mciteBstWouldAddEndPuncttrue
\mciteSetBstMidEndSepPunct{\mcitedefaultmidpunct}
{\mcitedefaultendpunct}{\mcitedefaultseppunct}\relax
\EndOfBibitem
\bibitem[Vdovin \latin{et~al.}(2009)Vdovin, Waluk, Dick, and Slenczka]{VdovinSlenczka2009}
Vdovin,~A.; Waluk,~J.; Dick,~B.; Slenczka,~A. Mode-Selective Promotion and Isotope Effects of Concerted Double-Hydrogen Tunneling in Porphycene Embedded in Superfluid Helium Nanodroplets. \emph{ChemPhysChem} \textbf{2009}, \emph{10}, 761--765\relax
\mciteBstWouldAddEndPuncttrue
\mciteSetBstMidEndSepPunct{\mcitedefaultmidpunct}
{\mcitedefaultendpunct}{\mcitedefaultseppunct}\relax
\EndOfBibitem
\bibitem[Talbot \latin{et~al.}(2020)Talbot, Yang, Huang, Duong, McCoy, Steele, and Johnson]{TalbotJohnson2020}
Talbot,~J.~J.; Yang,~N.; Huang,~M.; Duong,~C.~H.; McCoy,~A.~B.; Steele,~R.~P.; Johnson,~M.~A. Spectroscopic Signatures of Mode-Dependent Tunnel Splitting in the Iodide–Water Binary Complex. \emph{J. Phys. Chem. A.} \textbf{2020}, \emph{124}, 2991--3001\relax
\mciteBstWouldAddEndPuncttrue
\mciteSetBstMidEndSepPunct{\mcitedefaultmidpunct}
{\mcitedefaultendpunct}{\mcitedefaultseppunct}\relax
\EndOfBibitem
\bibitem[Sen and Kohen(2010)Sen, and Kohen]{SenKohen2010}
Sen,~A.; Kohen,~A. Enzymatic tunneling and kinetic isotope effects: chemistry at the crossroads. \emph{J. Phys. Org. Chem.} \textbf{2010}, \emph{23}, 613--619\relax
\mciteBstWouldAddEndPuncttrue
\mciteSetBstMidEndSepPunct{\mcitedefaultmidpunct}
{\mcitedefaultendpunct}{\mcitedefaultseppunct}\relax
\EndOfBibitem
\bibitem[Hama \latin{et~al.}(2015)Hama, Ueta, Kouchi, and Watanabe]{HamaWatanabe2015}
Hama,~T.; Ueta,~H.; Kouchi,~A.; Watanabe,~N. Quantum tunneling observed without its characteristic large kinetic isotope effects. \emph{Proc. Natl. Acad. Sci. U.S.A.} \textbf{2015}, \emph{112}, 7438--7443\relax
\mciteBstWouldAddEndPuncttrue
\mciteSetBstMidEndSepPunct{\mcitedefaultmidpunct}
{\mcitedefaultendpunct}{\mcitedefaultseppunct}\relax
\EndOfBibitem
\bibitem[Klinman and Offenbacher(2018)Klinman, and Offenbacher]{KlinmanOffenbacher2018}
Klinman,~J.~P.; Offenbacher,~A.~R. Understanding Biological Hydrogen Transfer Through the Lens of Temperature Dependent Kinetic Isotope Effects. \emph{Acc. Chem. Res.} \textbf{2018}, \emph{51}, 1966--1974\relax
\mciteBstWouldAddEndPuncttrue
\mciteSetBstMidEndSepPunct{\mcitedefaultmidpunct}
{\mcitedefaultendpunct}{\mcitedefaultseppunct}\relax
\EndOfBibitem
\bibitem[Hammes-Schiffer(2025)]{SHS2025}
Hammes-Schiffer,~S. Explaining Kinetic Isotope Effects in Proton-Coupled Electron Transfer Reactions. \emph{Acc. Chem. Res.} \textbf{2025}, \emph{58}, 1335--1344\relax
\mciteBstWouldAddEndPuncttrue
\mciteSetBstMidEndSepPunct{\mcitedefaultmidpunct}
{\mcitedefaultendpunct}{\mcitedefaultseppunct}\relax
\EndOfBibitem
\bibitem[Mil’nikov and Nakamura(2001)Mil’nikov, and Nakamura]{Mil'nikoovNakamura2001}
Mil’nikov,~G.~V.; Nakamura,~H. Practical implementation of the instanton theory for the ground-state tunneling splitting. \emph{J. Chem. Phys.} \textbf{2001}, \emph{115}, 6881--6897\relax
\mciteBstWouldAddEndPuncttrue
\mciteSetBstMidEndSepPunct{\mcitedefaultmidpunct}
{\mcitedefaultendpunct}{\mcitedefaultseppunct}\relax
\EndOfBibitem
\bibitem[Richardson and Althorpe(2011)Richardson, and Althorpe]{RichardsonAlthorpe2011}
Richardson,~J.~O.; Althorpe,~S.~C. Ring-polymer instanton method for calculating tunneling splittings. \emph{J. Chem. Phys.} \textbf{2011}, \emph{134}, 054109\relax
\mciteBstWouldAddEndPuncttrue
\mciteSetBstMidEndSepPunct{\mcitedefaultmidpunct}
{\mcitedefaultendpunct}{\mcitedefaultseppunct}\relax
\EndOfBibitem
\bibitem[Trenins \latin{et~al.}(2023)Trenins, Meuser, Bertschi, Vavourakis, Flütsch, and Richardson]{TreninsMeuserBertschiVavourakisFlutschRichardson2023}
Trenins,~G.; Meuser,~L.; Bertschi,~H.; Vavourakis,~O.; Flütsch,~R.; Richardson,~J.~O. Exact tunneling splittings from symmetrized path integrals. \emph{J. Chem. Phys.} \textbf{2023}, \emph{159}, 034108\relax
\mciteBstWouldAddEndPuncttrue
\mciteSetBstMidEndSepPunct{\mcitedefaultmidpunct}
{\mcitedefaultendpunct}{\mcitedefaultseppunct}\relax
\EndOfBibitem
\bibitem[Gregory and Clary(1995)Gregory, and Clary]{GregoryClary1995}
Gregory,~J.~K.; Clary,~D.~C. Calculations of the tunneling splittings in water dimer and trimer using diffusion Monte Carlo. \emph{J. Chem. Phys.} \textbf{1995}, \emph{102}, 7817--7829\relax
\mciteBstWouldAddEndPuncttrue
\mciteSetBstMidEndSepPunct{\mcitedefaultmidpunct}
{\mcitedefaultendpunct}{\mcitedefaultseppunct}\relax
\EndOfBibitem
\bibitem[Fortenberry \latin{et~al.}(2015)Fortenberry, Yu, Mancini, Bowman, Lee, Crawford, Klemperer, and Francisco]{FortenberryFrancisco2015}
Fortenberry,~R.~C.; Yu,~Q.; Mancini,~J.~S.; Bowman,~J.~M.; Lee,~T.~J.; Crawford,~T.~D.; Klemperer,~W.~F.; Francisco,~J.~S. Communication: Spectroscopic consequences of proton delocalization in OCHCO+. \emph{J. Chem. Phys.} \textbf{2015}, \emph{143}, 071102\relax
\mciteBstWouldAddEndPuncttrue
\mciteSetBstMidEndSepPunct{\mcitedefaultmidpunct}
{\mcitedefaultendpunct}{\mcitedefaultseppunct}\relax
\EndOfBibitem
\bibitem[DiRisio \latin{et~al.}(2022)DiRisio, Finney, and McCoy]{DiRisioFinneyMcCoy2022}
DiRisio,~R.~J.; Finney,~J.~M.; McCoy,~A.~B. Diffusion Monte Carlo approaches for studying nuclear quantum effects in fluxional molecules. \emph{Wiley Interdiscip. Rev. Comput. Mol. Sci.} \textbf{2022}, \emph{12}, e1615\relax
\mciteBstWouldAddEndPuncttrue
\mciteSetBstMidEndSepPunct{\mcitedefaultmidpunct}
{\mcitedefaultendpunct}{\mcitedefaultseppunct}\relax
\EndOfBibitem
\bibitem[Finney and McCoy(2024)Finney, and McCoy]{FinneyMcCoy2024}
Finney,~J.~M.; McCoy,~A.~B. Correlations between the Structures and Spectra of Protonated Water Clusters. \emph{J. Phys. Chem. A} \textbf{2024}, \emph{128}, 868--879\relax
\mciteBstWouldAddEndPuncttrue
\mciteSetBstMidEndSepPunct{\mcitedefaultmidpunct}
{\mcitedefaultendpunct}{\mcitedefaultseppunct}\relax
\EndOfBibitem
\bibitem[Coutinho-Neto \latin{et~al.}(2004)Coutinho-Neto, Viel, and Manthe]{Coutinho-NetoVielManthe2004}
Coutinho-Neto,~M.~D.; Viel,~A.; Manthe,~U. The ground state tunneling splitting of malonaldehyde: Accurate full dimensional quantum dynamics calculations. \emph{J. Chem. Phys.} \textbf{2004}, \emph{121}, 9207--9210\relax
\mciteBstWouldAddEndPuncttrue
\mciteSetBstMidEndSepPunct{\mcitedefaultmidpunct}
{\mcitedefaultendpunct}{\mcitedefaultseppunct}\relax
\EndOfBibitem
\bibitem[Schröder \latin{et~al.}(2011)Schröder, Gatti, and Meyer]{SchröderGattiMeyer2011}
Schröder,~M.; Gatti,~F.; Meyer,~H.-D. Theoretical studies of the tunneling splitting of malonaldehyde using the multiconfiguration time-dependent Hartree approach. \emph{J. Chem. Phys.} \textbf{2011}, \emph{134}, 234307\relax
\mciteBstWouldAddEndPuncttrue
\mciteSetBstMidEndSepPunct{\mcitedefaultmidpunct}
{\mcitedefaultendpunct}{\mcitedefaultseppunct}\relax
\EndOfBibitem
\bibitem[Suzuki \latin{et~al.}(2023)Suzuki, Kanno, Koseki, and Kono]{SuzukiKono2023}
Suzuki,~K.; Kanno,~M.; Koseki,~S.; Kono,~H. A Structure-Based Gaussian Expansion for Quantum Reaction Dynamics in Molecules: Application to Hydrogen Tunneling in Malonaldehyde. \emph{J. Phys. Chem. A.} \textbf{2023}, \emph{127}, 4152--4165\relax
\mciteBstWouldAddEndPuncttrue
\mciteSetBstMidEndSepPunct{\mcitedefaultmidpunct}
{\mcitedefaultendpunct}{\mcitedefaultseppunct}\relax
\EndOfBibitem
\bibitem[Webb \latin{et~al.}(2002)Webb, Iordanov, and Hammes-Schiffer]{WebbIordanovSHS2002}
Webb,~S.~P.; Iordanov,~T.; Hammes-Schiffer,~S. Multiconfigurational nuclear-electronic orbital approach: Incorporation of nuclear quantum effects in electronic structure calculations. \emph{J. Chem. Phys.} \textbf{2002}, \emph{117}, 4106--4118\relax
\mciteBstWouldAddEndPuncttrue
\mciteSetBstMidEndSepPunct{\mcitedefaultmidpunct}
{\mcitedefaultendpunct}{\mcitedefaultseppunct}\relax
\EndOfBibitem
\bibitem[Hammes-Schiffer(2021)]{SHS2021}
Hammes-Schiffer,~S. Nuclear–electronic orbital methods: Foundations and prospects. \emph{J. Chem. Phys.} \textbf{2021}, \emph{155}, 030901\relax
\mciteBstWouldAddEndPuncttrue
\mciteSetBstMidEndSepPunct{\mcitedefaultmidpunct}
{\mcitedefaultendpunct}{\mcitedefaultseppunct}\relax
\EndOfBibitem
\bibitem[Pak \latin{et~al.}(2007)Pak, Chakraborty, and Hammes-Schiffer]{PakChakrabortySHS2007}
Pak,~M.~V.; Chakraborty,~A.; Hammes-Schiffer,~S. Density Functional Theory Treatment of Electron Correlation in the Nuclear-Electronic Orbital Approach. \emph{J. Phys. Chem. A} \textbf{2007}, \emph{111}, 4522--4526\relax
\mciteBstWouldAddEndPuncttrue
\mciteSetBstMidEndSepPunct{\mcitedefaultmidpunct}
{\mcitedefaultendpunct}{\mcitedefaultseppunct}\relax
\EndOfBibitem
\bibitem[Yang \latin{et~al.}(2017)Yang, Brorsen, Culpitt, Pak, and Hammes-Schiffer]{YangSHS2017}
Yang,~Y.; Brorsen,~K.~R.; Culpitt,~T.; Pak,~M.~V.; Hammes-Schiffer,~S. Development of a practical multicomponent density functional for electron-proton correlation to produce accurate proton densities. \emph{J. Chem. Phys.} \textbf{2017}, \emph{147}, 114113\relax
\mciteBstWouldAddEndPuncttrue
\mciteSetBstMidEndSepPunct{\mcitedefaultmidpunct}
{\mcitedefaultendpunct}{\mcitedefaultseppunct}\relax
\EndOfBibitem
\bibitem[Brorsen \latin{et~al.}(2017)Brorsen, Yang, and Hammes-Schiffer]{BrorsenYangSHS2017}
Brorsen,~K.~R.; Yang,~Y.; Hammes-Schiffer,~S. Multicomponent Density Functional Theory: Impact of Nuclear Quantum Effects on Proton Affinities and Geometries. \emph{J. Phys. Chem. Lett.} \textbf{2017}, \emph{8}, 3488--3493\relax
\mciteBstWouldAddEndPuncttrue
\mciteSetBstMidEndSepPunct{\mcitedefaultmidpunct}
{\mcitedefaultendpunct}{\mcitedefaultseppunct}\relax
\EndOfBibitem
\bibitem[Yang \latin{et~al.}(2018)Yang, Culpitt, and Hammes-Schiffer]{YangCulpittSHS2018}
Yang,~Y.; Culpitt,~T.; Hammes-Schiffer,~S. Multicomponent Time-Dependent Density Functional Theory: Proton and Electron Excitation Energies. \emph{J. Phys. Chem. Lett.} \textbf{2018}, \emph{9}, 1765--1770\relax
\mciteBstWouldAddEndPuncttrue
\mciteSetBstMidEndSepPunct{\mcitedefaultmidpunct}
{\mcitedefaultendpunct}{\mcitedefaultseppunct}\relax
\EndOfBibitem
\bibitem[Culpitt \latin{et~al.}(2019)Culpitt, Yang, Pavošević, Tao, and Hammes-Schiffer]{CulpittSHS2019}
Culpitt,~T.; Yang,~Y.; Pavošević,~F.; Tao,~Z.; Hammes-Schiffer,~S. Enhancing the applicability of multicomponent time-dependent density functional theory. \emph{J. Chem. Phys.} \textbf{2019}, \emph{150}, 201101\relax
\mciteBstWouldAddEndPuncttrue
\mciteSetBstMidEndSepPunct{\mcitedefaultmidpunct}
{\mcitedefaultendpunct}{\mcitedefaultseppunct}\relax
\EndOfBibitem
\bibitem[Pavošević \latin{et~al.}(2019)Pavošević, Culpitt, and Hammes-Schiffer]{PavoševićCulpittSHS2019}
Pavošević,~F.; Culpitt,~T.; Hammes-Schiffer,~S. Multicomponent Coupled Cluster Singles and Doubles Theory within the Nuclear-Electronic Orbital Framework. \emph{J. of Chem. Theory Comput.} \textbf{2019}, \emph{15}, 338--347\relax
\mciteBstWouldAddEndPuncttrue
\mciteSetBstMidEndSepPunct{\mcitedefaultmidpunct}
{\mcitedefaultendpunct}{\mcitedefaultseppunct}\relax
\EndOfBibitem
\bibitem[Pavošević \latin{et~al.}(2021)Pavošević, Tao, and Hammes-Schiffer]{PavoševićTaoSHS2021}
Pavošević,~F.; Tao,~Z.; Hammes-Schiffer,~S. Multicomponent Coupled Cluster Singles and Doubles with Density Fitting: Protonated Water Tetramers with Quantized Protons. \emph{J. Chem. Phys. Lett.} \textbf{2021}, \emph{12}, 1631--1637\relax
\mciteBstWouldAddEndPuncttrue
\mciteSetBstMidEndSepPunct{\mcitedefaultmidpunct}
{\mcitedefaultendpunct}{\mcitedefaultseppunct}\relax
\EndOfBibitem
\bibitem[Pavošević and Hammes-Schiffer(2022)Pavošević, and Hammes-Schiffer]{PavoševićSHS2022}
Pavošević,~F.; Hammes-Schiffer,~S. Triple electron–electron–proton excitations and second-order approximations in nuclear–electronic orbital coupled cluster methods. \emph{J. of Chem. Phys.} \textbf{2022}, \emph{157}, 074104\relax
\mciteBstWouldAddEndPuncttrue
\mciteSetBstMidEndSepPunct{\mcitedefaultmidpunct}
{\mcitedefaultendpunct}{\mcitedefaultseppunct}\relax
\EndOfBibitem
\bibitem[Muolo \latin{et~al.}(2020)Muolo, Baiardi, Feldmann, and Reiher]{MuoloBaiardiFeldmannReiher2020}
Muolo,~A.; Baiardi,~A.; Feldmann,~R.; Reiher,~M. Nuclear-electronic all-particle density matrix renormalization group. \emph{The Journal of Chemical Physics} \textbf{2020}, \emph{152}, 204103\relax
\mciteBstWouldAddEndPuncttrue
\mciteSetBstMidEndSepPunct{\mcitedefaultmidpunct}
{\mcitedefaultendpunct}{\mcitedefaultseppunct}\relax
\EndOfBibitem
\bibitem[Feldmann \latin{et~al.}(2022)Feldmann, Muolo, Baiardi, and Reiher]{FeldmannMuoloBaiardiReiher2022}
Feldmann,~R.; Muolo,~A.; Baiardi,~A.; Reiher,~M. Quantum Proton Effects from Density Matrix Renormalization Group Calculations. \emph{J. Chem. Theory Comput.} \textbf{2022}, \emph{18}, 234--250\relax
\mciteBstWouldAddEndPuncttrue
\mciteSetBstMidEndSepPunct{\mcitedefaultmidpunct}
{\mcitedefaultendpunct}{\mcitedefaultseppunct}\relax
\EndOfBibitem
\bibitem[Fajen and Brorsen(2021)Fajen, and Brorsen]{FajenBrorsen2021}
Fajen,~O.~J.; Brorsen,~K.~R. Multicomponent CASSCF Revisited: Large Active Spaces Are Needed for Qualitatively Accurate Protonic Densities. \emph{J. Chem. Theory Comput.} \textbf{2021}, \emph{17}, 965--974\relax
\mciteBstWouldAddEndPuncttrue
\mciteSetBstMidEndSepPunct{\mcitedefaultmidpunct}
{\mcitedefaultendpunct}{\mcitedefaultseppunct}\relax
\EndOfBibitem
\bibitem[Pak and Hammes-Schiffer(2004)Pak, and Hammes-Schiffer]{PakSHS2004}
Pak,~M.~V.; Hammes-Schiffer,~S. Electron-Proton Correlation for Hydrogen Tunneling Systems. \emph{Phys. Rev. Lett.} \textbf{2004}, \emph{92}, 103002\relax
\mciteBstWouldAddEndPuncttrue
\mciteSetBstMidEndSepPunct{\mcitedefaultmidpunct}
{\mcitedefaultendpunct}{\mcitedefaultseppunct}\relax
\EndOfBibitem
\bibitem[Pak \latin{et~al.}(2004)Pak, Swalina, Webb, and Hammes-Schiffer]{PakSwalinaWebbSHS2004}
Pak,~M.~V.; Swalina,~C.; Webb,~S.~P.; Hammes-Schiffer,~S. Application of the nuclear–electronic orbital method to hydrogen transfer systems: multiple centers and multiconfigurational wavefunctions. \emph{Chem. Phys.} \textbf{2004}, \emph{304}, 227--236\relax
\mciteBstWouldAddEndPuncttrue
\mciteSetBstMidEndSepPunct{\mcitedefaultmidpunct}
{\mcitedefaultendpunct}{\mcitedefaultseppunct}\relax
\EndOfBibitem
\bibitem[Swalina \latin{et~al.}(2005)Swalina, Pak, and Hammes-Schiffer]{SwalinaPakSHS2005}
Swalina,~C.; Pak,~M.~V.; Hammes-Schiffer,~S. Analysis of the nuclear-electronic orbital method for model hydrogen transfer systems. \emph{J. Chem. Phys.} \textbf{2005}, \emph{123}, 014303\relax
\mciteBstWouldAddEndPuncttrue
\mciteSetBstMidEndSepPunct{\mcitedefaultmidpunct}
{\mcitedefaultendpunct}{\mcitedefaultseppunct}\relax
\EndOfBibitem
\bibitem[Skone \latin{et~al.}(2005)Skone, Pak, and Hammes-Schiffer]{SkonePakSHS2005}
Skone,~J.~H.; Pak,~M.~V.; Hammes-Schiffer,~S. Nuclear-electronic orbital nonorthogonal configuration interaction approach. \emph{J. Chem. Phys.} \textbf{2005}, \emph{123}, 134108\relax
\mciteBstWouldAddEndPuncttrue
\mciteSetBstMidEndSepPunct{\mcitedefaultmidpunct}
{\mcitedefaultendpunct}{\mcitedefaultseppunct}\relax
\EndOfBibitem
\bibitem[Yu and Hammes-Schiffer(2020)Yu, and Hammes-Schiffer]{YuSHS2020}
Yu,~Q.; Hammes-Schiffer,~S. Nuclear-Electronic Orbital Multistate Density Functional Theory. \emph{J. Phys. Chem. Lett.} \textbf{2020}, \emph{11}, 10106--10113\relax
\mciteBstWouldAddEndPuncttrue
\mciteSetBstMidEndSepPunct{\mcitedefaultmidpunct}
{\mcitedefaultendpunct}{\mcitedefaultseppunct}\relax
\EndOfBibitem
\bibitem[Dickinson \latin{et~al.}(2023)Dickinson, Yu, and Hammes-Schiffer]{DickinsonYuSHS2023}
Dickinson,~J.~A.; Yu,~Q.; Hammes-Schiffer,~S. Generalized Nuclear-Electronic Orbital Multistate Density Functional Theory for Multiple Proton Transfer Processes. \emph{J. Phys. Chem. Lett.} \textbf{2023}, \emph{14}, 6170--6178\relax
\mciteBstWouldAddEndPuncttrue
\mciteSetBstMidEndSepPunct{\mcitedefaultmidpunct}
{\mcitedefaultendpunct}{\mcitedefaultseppunct}\relax
\EndOfBibitem
\bibitem[Garner \latin{et~al.}(2024)Garner, Upadhyay, Li, and Hammes-Schiffer]{GarnerUpadhyayLiSHS2024}
Garner,~S.~M.; Upadhyay,~S.; Li,~X.; Hammes-Schiffer,~S. Nuclear–Electronic Orbital Time-Dependent Configuration Interaction Method. \emph{J. Phys. Chem. Lett.} \textbf{2024}, \emph{15}, 6017--6023\relax
\mciteBstWouldAddEndPuncttrue
\mciteSetBstMidEndSepPunct{\mcitedefaultmidpunct}
{\mcitedefaultendpunct}{\mcitedefaultseppunct}\relax
\EndOfBibitem
\bibitem[Yu \latin{et~al.}(2022)Yu, Roy, and Hammes-Schiffer]{YuRoySHS2022}
Yu,~Q.; Roy,~S.; Hammes-Schiffer,~S. Nonadiabatic Dynamics of Hydrogen Tunneling with Nuclear-Electronic Orbital Multistate Density Functional Theory. \emph{J. Chem. Theory Comput.} \textbf{2022}, \emph{18}, 7132--7141\relax
\mciteBstWouldAddEndPuncttrue
\mciteSetBstMidEndSepPunct{\mcitedefaultmidpunct}
{\mcitedefaultendpunct}{\mcitedefaultseppunct}\relax
\EndOfBibitem
\bibitem[Dickinson and Hammes-Schiffer(2024)Dickinson, and Hammes-Schiffer]{DickinsonSHS2024}
Dickinson,~J.~A.; Hammes-Schiffer,~S. Nonadiabatic Hydrogen Tunneling Dynamics for Multiple Proton Transfer Processes with Generalized Nuclear-Electronic Orbital Multistate Density Functional Theory. \emph{J. Chem. Theory Comput.} \textbf{2024}, \emph{20}, 7716--7727\relax
\mciteBstWouldAddEndPuncttrue
\mciteSetBstMidEndSepPunct{\mcitedefaultmidpunct}
{\mcitedefaultendpunct}{\mcitedefaultseppunct}\relax
\EndOfBibitem
\bibitem[Malbon and Hammes-Schiffer(2025)Malbon, and Hammes-Schiffer]{MalbonSHS2025}
Malbon,~C.~L.; Hammes-Schiffer,~S. Nuclear-Electronic Orbital Multireference Configuration Interaction for Ground and Excited Vibronic States and Fundamental Insights into Multicomponent Basis Sets. \emph{Journal of Chemical Theory and Computation} \textbf{2025}, \emph{21}, 3968--3980\relax
\mciteBstWouldAddEndPuncttrue
\mciteSetBstMidEndSepPunct{\mcitedefaultmidpunct}
{\mcitedefaultendpunct}{\mcitedefaultseppunct}\relax
\EndOfBibitem
\bibitem[Lischka \latin{et~al.}(2018)Lischka, Nachtigallová, Aquino, Szalay, Plasser, Machado, and Barbatti]{Lishchka2018}
Lischka,~H.; Nachtigallová,~D.; Aquino,~A. J.~A.; Szalay,~P.~G.; Plasser,~F.; Machado,~F. B.~C.; Barbatti,~M. Multireference Approaches for Excited States of Molecules. \emph{Chem. Rev.} \textbf{2018}, \emph{118}, 7293--7361\relax
\mciteBstWouldAddEndPuncttrue
\mciteSetBstMidEndSepPunct{\mcitedefaultmidpunct}
{\mcitedefaultendpunct}{\mcitedefaultseppunct}\relax
\EndOfBibitem
\bibitem[Matsika(2021)]{Matsika2021}
Matsika,~S. Electronic Structure Methods for the Description of Nonadiabatic Effects and Conical Intersections. \emph{Chem. Rev.} \textbf{2021}, \emph{121}, 9407--9449\relax
\mciteBstWouldAddEndPuncttrue
\mciteSetBstMidEndSepPunct{\mcitedefaultmidpunct}
{\mcitedefaultendpunct}{\mcitedefaultseppunct}\relax
\EndOfBibitem
\bibitem[Terrill and Nesbitt(2010)Terrill, and Nesbitt]{TerrillNesbitt2010}
Terrill,~K.; Nesbitt,~D.~J. Ab initio anharmonic vibrational frequency predictions for linear proton-bound complexes OC–H+–CO and N2–H+–N2. \emph{Phys. Chem. Chem. Phys.} \textbf{2010}, \emph{12}, 8311--8322\relax
\mciteBstWouldAddEndPuncttrue
\mciteSetBstMidEndSepPunct{\mcitedefaultmidpunct}
{\mcitedefaultendpunct}{\mcitedefaultseppunct}\relax
\EndOfBibitem
\bibitem[Shida \latin{et~al.}(1989)Shida, Barbara, and Almlöf]{ShidaBarbaraAlmlöf1989}
Shida,~N.; Barbara,~P.~F.; Almlöf,~J.~E. A theoretical study of multidimensional nuclear tunneling in malonaldehyde. \emph{J. Chem. Phys.} \textbf{1989}, \emph{91}, 4061--4072\relax
\mciteBstWouldAddEndPuncttrue
\mciteSetBstMidEndSepPunct{\mcitedefaultmidpunct}
{\mcitedefaultendpunct}{\mcitedefaultseppunct}\relax
\EndOfBibitem
\bibitem[Mizukami \latin{et~al.}(2014)Mizukami, Habershon, and Tew]{MizukamiHabershonTew2014}
Mizukami,~W.; Habershon,~S.; Tew,~D.~P. A compact and accurate semi-global potential energy surface for malonaldehyde from constrained least squares regression. \emph{J. Chem. Phys.} \textbf{2014}, \emph{141}, 144310\relax
\mciteBstWouldAddEndPuncttrue
\mciteSetBstMidEndSepPunct{\mcitedefaultmidpunct}
{\mcitedefaultendpunct}{\mcitedefaultseppunct}\relax
\EndOfBibitem
\bibitem[Vaillant \latin{et~al.}(2018)Vaillant, Wales, and Althorpe]{VaillantWalesAlthorpe2018}
Vaillant,~C.~L.; Wales,~D.~J.; Althorpe,~S.~C. Tunneling splittings from path-integral molecular dynamics using a Langevin thermostat. \emph{J. Chem. Phys.} \textbf{2018}, \emph{148}, 234102\relax
\mciteBstWouldAddEndPuncttrue
\mciteSetBstMidEndSepPunct{\mcitedefaultmidpunct}
{\mcitedefaultendpunct}{\mcitedefaultseppunct}\relax
\EndOfBibitem
\bibitem[Lawrence \latin{et~al.}(2023)Lawrence, Dušek, and Richardson]{LawrenceDusekRichardson2023}
Lawrence,~J.~E.; Dušek,~J.; Richardson,~J.~O. Perturbatively corrected ring-polymer instanton theory for accurate tunneling splittings. \emph{J. Chem. Phys.} \textbf{2023}, \emph{159}, 014111\relax
\mciteBstWouldAddEndPuncttrue
\mciteSetBstMidEndSepPunct{\mcitedefaultmidpunct}
{\mcitedefaultendpunct}{\mcitedefaultseppunct}\relax
\EndOfBibitem
\bibitem[Lauvergnat and Nauts(2023)Lauvergnat, and Nauts]{LauvergnatNauts2023}
Lauvergnat,~D.; Nauts,~A. Smolyak Scheme for solving the Schrödinger equation: Application to Malonaldehyde in Full Dimensionality. \emph{ChemPhysChem} \textbf{2023}, \emph{24}, e202300501\relax
\mciteBstWouldAddEndPuncttrue
\mciteSetBstMidEndSepPunct{\mcitedefaultmidpunct}
{\mcitedefaultendpunct}{\mcitedefaultseppunct}\relax
\EndOfBibitem
\bibitem[Baumann \latin{et~al.}(2025)Baumann, Trenins, and Richardson]{BaumannTreninsRichardson2025}
Baumann,~J.; Trenins,~G.; Richardson,~J.~O. The exact tunnelling splitting of malonaldehyde from symmetrized path-integral molecular dynamics. \emph{Mol. Phys.} \textbf{2025}, e2474202\relax
\mciteBstWouldAddEndPuncttrue
\mciteSetBstMidEndSepPunct{\mcitedefaultmidpunct}
{\mcitedefaultendpunct}{\mcitedefaultseppunct}\relax
\EndOfBibitem
\bibitem[Marston and Balint‐Kurti(1989)Marston, and Balint‐Kurti]{MarstonBalintKurti1989}
Marston,~C.~C.; Balint‐Kurti,~G.~G. The Fourier grid Hamiltonian method for bound state eigenvalues and eigenfunctions. \emph{J. Chem. Phys.} \textbf{1989}, \emph{91}, 3571--3576\relax
\mciteBstWouldAddEndPuncttrue
\mciteSetBstMidEndSepPunct{\mcitedefaultmidpunct}
{\mcitedefaultendpunct}{\mcitedefaultseppunct}\relax
\EndOfBibitem
\bibitem[Webb and Hammes-Schiffer(2000)Webb, and Hammes-Schiffer]{WebbSHS2000}
Webb,~S.~P.; Hammes-Schiffer,~S. Fourier grid Hamiltonian multiconfigurational self-consistent-field: A method to calculate multidimensional hydrogen vibrational wavefunctions. \emph{J. Chem. Phys.} \textbf{2000}, \emph{113}, 5214--5227\relax
\mciteBstWouldAddEndPuncttrue
\mciteSetBstMidEndSepPunct{\mcitedefaultmidpunct}
{\mcitedefaultendpunct}{\mcitedefaultseppunct}\relax
\EndOfBibitem
\bibitem[Kendall \latin{et~al.}(1992)Kendall, Dunning, and Harrison]{KendallDunningHarrison1992}
Kendall,~R.~A.; Dunning,~J.,~Thom~H.; Harrison,~R.~J. Electron affinities of the first‐row atoms revisited. Systematic basis sets and wave functions. \emph{J. Chem. Phys.} \textbf{1992}, \emph{96}, 6796--6806\relax
\mciteBstWouldAddEndPuncttrue
\mciteSetBstMidEndSepPunct{\mcitedefaultmidpunct}
{\mcitedefaultendpunct}{\mcitedefaultseppunct}\relax
\EndOfBibitem
\bibitem[Dunning(1989)]{Dunning1989}
Dunning,~J.,~Thom~H. Gaussian basis sets for use in correlated molecular calculations. I. The atoms boron through neon and hydrogen. \emph{J. Chem. Phys.} \textbf{1989}, \emph{90}, 1007--1023\relax
\mciteBstWouldAddEndPuncttrue
\mciteSetBstMidEndSepPunct{\mcitedefaultmidpunct}
{\mcitedefaultendpunct}{\mcitedefaultseppunct}\relax
\EndOfBibitem
\bibitem[Hazra \latin{et~al.}(2009)Hazra, Skone, and Hammes-Schiffer]{HazraSkoneSHS2009}
Hazra,~A.; Skone,~J.~H.; Hammes-Schiffer,~S. Combining the nuclear-electronic orbital approach with vibronic coupling theory: Calculation of the tunneling splitting for malonaldehyde. \emph{J. Chem. Phys.} \textbf{2009}, \emph{130}, 054108\relax
\mciteBstWouldAddEndPuncttrue
\mciteSetBstMidEndSepPunct{\mcitedefaultmidpunct}
{\mcitedefaultendpunct}{\mcitedefaultseppunct}\relax
\EndOfBibitem
\bibitem[Helgaker \latin{et~al.}(2000)Helgaker, Jørgensen, and Olsen]{Helgaker2000}
Helgaker,~T.; Jørgensen,~P.; Olsen,~J. \emph{Molecular Electronic‐Structure Theory}; John Wiley \& Sons, Ltd, 2000; Chapter 8, pp 287--335\relax
\mciteBstWouldAddEndPuncttrue
\mciteSetBstMidEndSepPunct{\mcitedefaultmidpunct}
{\mcitedefaultendpunct}{\mcitedefaultseppunct}\relax
\EndOfBibitem
\bibitem[Tautermann \latin{et~al.}(2002)Tautermann, Voegele, Loerting, and Liedl]{TautermannLiedl2002}
Tautermann,~C.~S.; Voegele,~A.~F.; Loerting,~T.; Liedl,~K.~R. The optimal tunneling path for the proton transfer in malonaldehyde. \emph{J. Chem. Phys.} \textbf{2002}, \emph{117}, 1962--1966\relax
\mciteBstWouldAddEndPuncttrue
\mciteSetBstMidEndSepPunct{\mcitedefaultmidpunct}
{\mcitedefaultendpunct}{\mcitedefaultseppunct}\relax
\EndOfBibitem
\end{mcitethebibliography}

\end{document}


\newpage

\tableofcontents

\newpage

\section{Tunneling Splitting Tables}

\begin{table}[H]
\let\TPToverlap=\TPTrlap
\centering
\caption{\ce{HeHHe+} Tunneling Splittings (in \si{\wn}) with Hydrogen and Deuterium Computed with NEO-MR-SD$_\text{en}$CI Compared to FGH Benchmark}
\label{table:hehhe_tun_split}
\begin{threeparttable}
\begin{tabularx}{1\textwidth}{Y Y Y Y Y Y}
\toprule
\toprule
\multicolumn{1}{c}{He---He (\si{\angstrom})} & \multicolumn{3}{c}{NEO-MR-SD$_\text{en}$CI\tnote{\emph{a}}} & \multicolumn{2}{c}{FGH\tnote{\emph{b}}} \\
\midrule
 & \underline{H (8s8p8d)} & \underline{H (8s8p8d8f)} & \underline{D (8s8p8d8f)} & \underline{  H  } & \underline{  D  } \\
2.20 & 227.8 & 191.6 & 48.3 & 206.5 & 64.1 \\
2.25 & 118.8 & 93.5 & 14.0 & 101.0 & 20.4\\
2.30 & 55.1 & 38.8 & 2.4 & 42.1 & 5.3\\
2.35 & 23.2 & 12.8 & 0.6 & 15.4 & 1.2 \\
2.40\tnote{\emph{c}} & 9.0 & 2.3 & NA & 4.9 & NA \\

\bottomrule
\end{tabularx}
\begin{tablenotes}\linespread{1}\footnotesize
    \item[\emph{a}] The basis function center positions are optimized at the CCSD/aug-cc-pVTZ level of theory. The inidcated protonic basis set and the cc-pV5Z* electronic basis set are used on the proton basis function centers. The aug-cc-pVTZ electronic basis set is used on He. No electronic active space and a full protonic active space are used.
    \item[\emph{b}] The FGH results were computed at the CCSD/aug-cc-pVTZ level of theory. 
                    The grid is 32$\times$32$\times$32 with a step size of 0.055 Å. The grid ranges from $-0.820$ to 0.875 Å in the $x$, $y$, and $z$ directions with the molecule centered at the origin.
    \item[\emph{c}] The deuterium splitting is too small to calculate reliably with these methods at this distance.
\end{tablenotes}
\end{threeparttable}
\end{table}

\begin{table}[H]
\let\TPToverlap=\TPTrlap
\centering
\caption{\ce{FHF-} Tunneling Splittings (in \si{\wn}) with Hydrogen and Deuterium Computed with NEO-MR-SD$_\text{en}$CI Compared to FGH Benchmark}
\label{table:fhf_tun_split}
\begin{threeparttable}
\begin{tabularx}{1\textwidth}{Y Y Y Y Y Y}
\toprule
\toprule
\multicolumn{1}{c}{F---F (\si{\angstrom})} & \multicolumn{3}{c}{NEO-MR-SD$_\text{en}$CI\tnote{\emph{a}}} & \multicolumn{2}{c}{FGH\tnote{\emph{b}}} \\
\midrule
 & \underline{H (8s8p8d)} & \underline{H (8s8p8d8f)} & \underline{D (8s8p8d8f)} & \underline{  H  } & \underline{  D  } \\
2.54 & 248.4 & 262.5 & 70.6 & 267.9 & 80.6 \\ 
2.58 & 127.3 & 141.3 & 25.6 & 143.3 & 28.7\\
2.62 & 57.9 & 66.4 & 7.7 & 66.0 & 8.4 \\
2.66 & 23.8 & 27.5 & 1.9 & 26.7 & 2.1 \\
2.70 & 6.2 & 10.1 & 0.4 & 9.7 & 0.5\\
2.74\tnote{\emph{c}} & 3.1 & 3.1 & NA & 3.2 & NA\\
\bottomrule
\end{tabularx}
\begin{tablenotes}\linespread{1}\footnotesize
    \item[\emph{a}] The basis function center positions are optimized at the CCSD/aug-cc-pVTZ level of theory. The indicated protonic basis set and the cc-pV5Z* electronic basis set are used on the proton basis function centers. The aug-cc-pVTZ electronic basis set is used on F. A (2e, 2o) electronic active space and a full protonic active space are used.
    \item[\emph{b}] The FGH results were computed at the CCSD/aug-cc-pVTZ level of theory.
                    The grid is 32$\times$32$\times$32 with a step size of 0.074 Å. The grid ranges from $-1.113$ to 1.187 Å in the $x$, $y$, and $z$ directions with the molecule centered at the origin.
    \item[\emph{c}] The deuterium splitting is too small to calculate reliably with these methods at this distance.
\end{tablenotes}
\end{threeparttable}
\end{table}

\begin{table}[H]
\let\TPToverlap=\TPTrlap
\centering
\caption{\ce{OCHCO+} Tunneling Splittings (in \si{\wn}) Computed with NEO-MR-SD$_\text{en}$CI Compared to FGH Benchmark}
\label{table:ochco_tun_split}
\begin{threeparttable}
\begin{tabularx}{1\textwidth}{Y Y Y Y}
\toprule
\toprule
C---C(Å)  & \multicolumn{2}{c}{NEO-MR-SD$_\text{en}$CI\tnote{\emph{a}}} & FGH\tnote{\emph{b}}\\
\midrule
& \underline{8s8p8d} & \underline{5s5p5d5f} & \\
2.90 & 196.4 & 229.2 & 224.3 \\
2.94 & 106.8 & 136.7 & 131.8 \\
2.98 & 53.8 & 74.5 & 69.1 \\
3.02 & 24.6 & 36.9 & 32.6 \\
3.06 & 10.3 & 16.8 & 14.1 \\
3.10 & 3.9 & 7.1 & 5.6 \\
\bottomrule
\end{tabularx}
\begin{tablenotes}\linespread{1}\footnotesize
    \item[\emph{a}] The basis function center positions are optimized at the CCSD/aug-cc-pVTZ level of theory. The indicated protonic basis set and the cc-pV5Z* electronic basis set are used on the proton basis function centers. The aug-cc-pVTZ electronic basis set is used on O and C. A (2e, 2o) electronic active space and a full protonic active space are used.
    \item[\emph{b}] The FGH results were computed at the CCSD/aug-cc-pVTZ level of theory.
                    The grid is 32$\times$32$\times$32 with a step size of 0.06 Å. The grid ranges from $-0.93$ to 0.93 Å in the $x$, $y$, and $z$ directions with the molecule centered at the origin.
\end{tablenotes}
\end{threeparttable}
\end{table}

\begin{table}[H]
\let\TPToverlap=\TPTrlap
\centering
\caption{Malonaldehyde Tunneling Splittings (in \si{\wn}) Computed with NEO-MR-SD$_\text{en}$CI Compared to FGH Benchmark}
\label{table:malon_tun_split}
\begin{threeparttable}
\begin{tabularx}{1\textwidth}{Y Y Y Y Y Y}
\toprule
\toprule
O---O(Å)  & \multicolumn{3}{c}{NEO-MR-SD$_\text{en}$CI\tnote{\emph{a}}} & \multicolumn{2}{c}{FGH\tnote{\emph{b}}}\\
\midrule
& \underline{H (8s8p8d)} & \underline{H (5s5p5d5f)} & \underline{D (5s5p5d5f)} & \underline{  H  } & \underline{  D  } \\
2.55 & 36.0 & 56.1 & 5.5  & 49.7 & 5.3\\
2.62 & 5.4 & 10.3 & 0.4  & 8.4 & 0.4\\
\bottomrule
\end{tabularx}
\begin{tablenotes}\linespread{1}\footnotesize
    \item[\emph{a}] The basis function center positions are optimized at the CCSD/aug-cc-pVTZ level of theory. The indicated protonic basis set and the cc-pV5Z* electronic basis set are used on the proton basis function centers. The aug-cc-pVTZ electronic basis set is used on the other atoms. A (2e, 2o) electronic active space and a full protonic active space are used.
    \item[\emph{b}] The FGH results were computed at the CCSD/cc-pVTZ level of theory.
                    The grid is 32$\times$32$\times$32 with a step size of 0.06 Å and ranges from $-0.93$ to 0.93 Å in the $x$ and $z$ directions for both geometries, as well as from $-1.253$ to 0.607 Å (from $-1.271$ to 0.589 A) in the $y$ direction for the O---O distance of 2.55 Å (2.62 Å) with the malonaldehyde coordinates given in Section 6.4.
\end{tablenotes}
\end{threeparttable}
\end{table}

\section{Basis Set Convergence}

\begin{table}[H]
\let\TPToverlap=\TPTrlap
\centering
\caption{\ce{HeHHe+} Tunneling Splittings (in \si{\wn}) Computed with NEO-MR-SD$_\text{en}$CI with Varying Protonic and Electronic Basis Sets\tnote{\emph{a}}}
\label{table:bas_set_conv_hehhe}
\begin{threeparttable}
\begin{tabularx}{1\textwidth}{Y Y Y Y Y}
\toprule
\toprule
Protonic Basis Set & Electronic Basis Set & NEO-MR-SD$_\text{en}$CI\tnote{\emph{b}}\\
\midrule
8s8p8d & cc-pV5Z* & 55.1 \\
8s8p8d8f & cc-pV5Z* & 38.8 \\
5s5p5d5f5g & cc-pV5Z* & 35.4 \\
\midrule
\midrule
8s8p8d8f & cc-pVQZ* & 45.6 \\
8s8p8d8f & cc-pV5Z* & 38.8\\
8s8p8d8f & cc-pV6Z* & 35.4\\
\midrule
\midrule
8s8p8d8f & cc-pV5Z* & 38.8\\
5s5p5d5f & cc-pV5Z* & 38.8\\
\bottomrule
\end{tabularx}
\begin{tablenotes}\linespread{1}\footnotesize
    \item[\emph{a}] The He---He is 2.3 Å, and the FGH tunneling splitting is 42.1 \si{\wn}.
    \item[\emph{b}] The basis function center positions are optimized at the CCSD/aug-cc-pVTZ level of theory. The indicated protonic and electronic basis sets are used on the proton basis function centers. The aug-cc-pVTZ electronic basis set is used on He. No electronic active space and a full protonic active space are used.
\end{tablenotes}
\end{threeparttable}
\end{table}

\begin{table}[H]
\let\TPToverlap=\TPTrlap
\centering
\caption{\ce{FHF-} Tunneling Splittings (in \si{\wn}) Computed with NEO-MR-SD$_\text{en}$CI with Varying Protonic Basis Sets\tnote{\emph{a}}}
\label{table:scaled_bas_deut}
\begin{threeparttable}
\begin{tabularx}{1\textwidth}{Y Y Y Y}
\toprule
\toprule
Protonic Basis Set & NEO-MR-SD$_\text{en}$CI\tnote{\emph{b}}\\
\midrule
8s8p8d & 23.8 \\
8s8p8d8f & 27.5 \\
5s5p5d5f5g & 28.9 \\
\bottomrule
\end{tabularx}
\begin{tablenotes}\linespread{1}\footnotesize
    \item[\emph{a}] The F---F distance is 2.66 Å, and the FGH tunneling splitting is 26.7 \si{\wn}.
    \item[\emph{b}] The basis function center positions are optimized at the CCSD/aug-cc-pVTZ level of theory. The indicated protonic basis set and the cc-pV5Z* electronic basis sets are used on the proton basis function centers. The aug-cc-pVTZ electronic basis set is used on F. A (2e,2o) electronic active space and a full protonic active space are used.
\end{tablenotes}
\end{threeparttable}
\end{table}

\newpage

\section{Electron Density}

\begin{figure}
    \centering
    \includegraphics[width=6in]{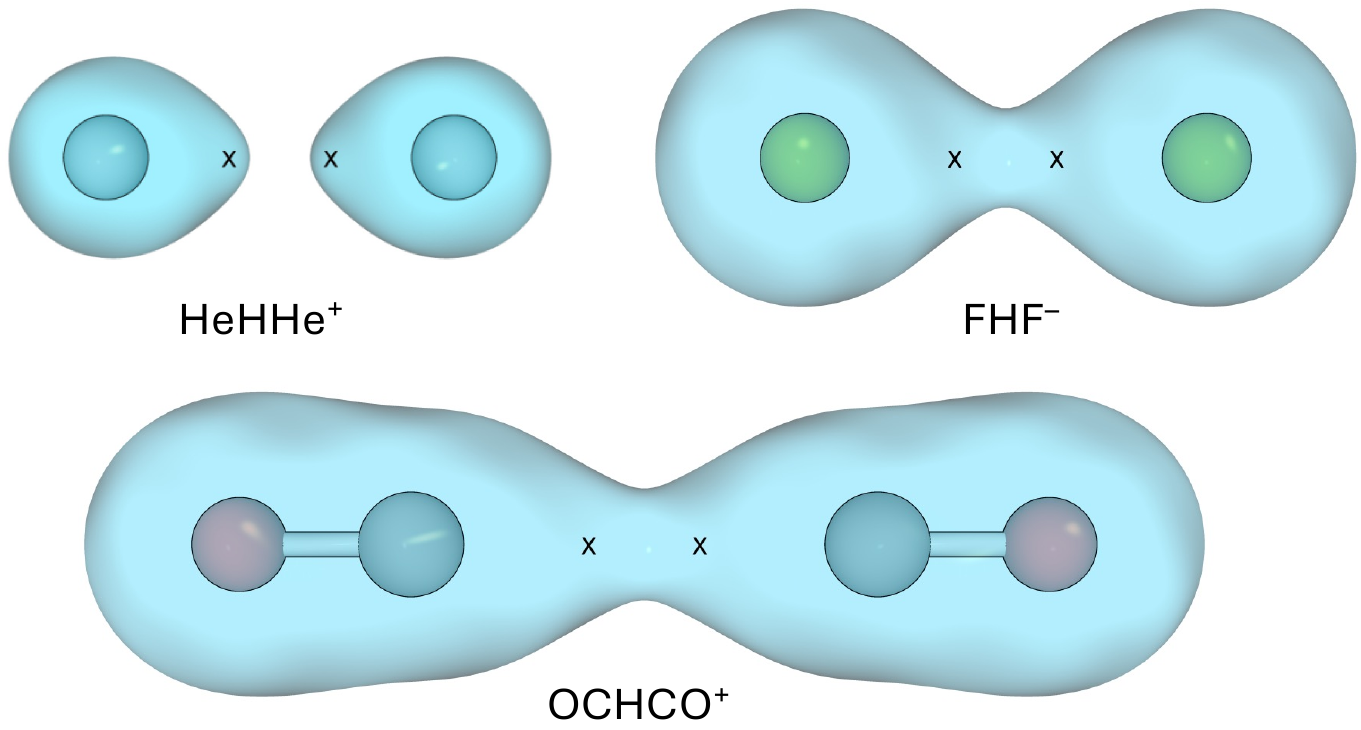}
    \caption{Electronic density of the ground state for \ce{HeHHe+}, \ce{OCHCO+}, and \ce{FHF-}. The isovalue of the electronic density is 0.04 e $\text{Bohr}^{-3}$ for all systems. The proton basis function center positions are indicated with an "X".}
    \label{fig:e_den}
\end{figure}

\newpage

\section{Electronic Active Spaces}

Protonic orbitals are not shown, as a full protonic active space was used.

\begin{figure}
    \centering
    \includegraphics[width=2in]{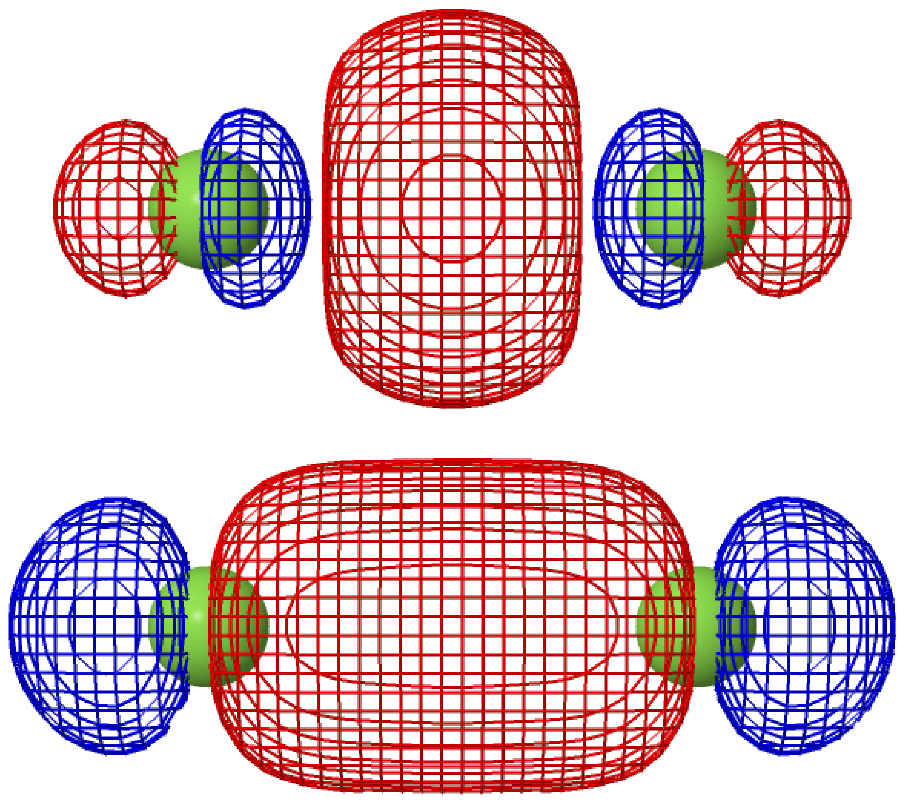}
    \caption{Electronic orbitals in the \ce{FHF-} active space.}
    \label{fig:fhf_act}
\end{figure}

\begin{figure}
    \centering
    \includegraphics[width=2in]{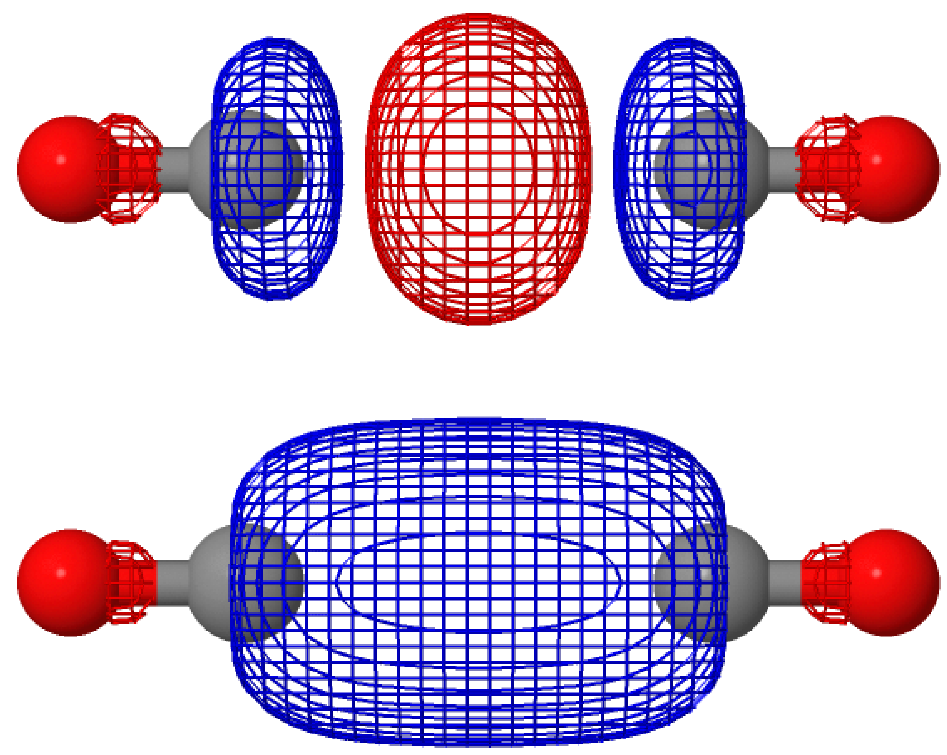}
    \caption{Electronic orbitals in the \ce{OCHCO+} active space.} 
    \label{fig:ochco_act}
\end{figure}

\begin{figure}
    \centering
    \includegraphics[width=0.5\linewidth]{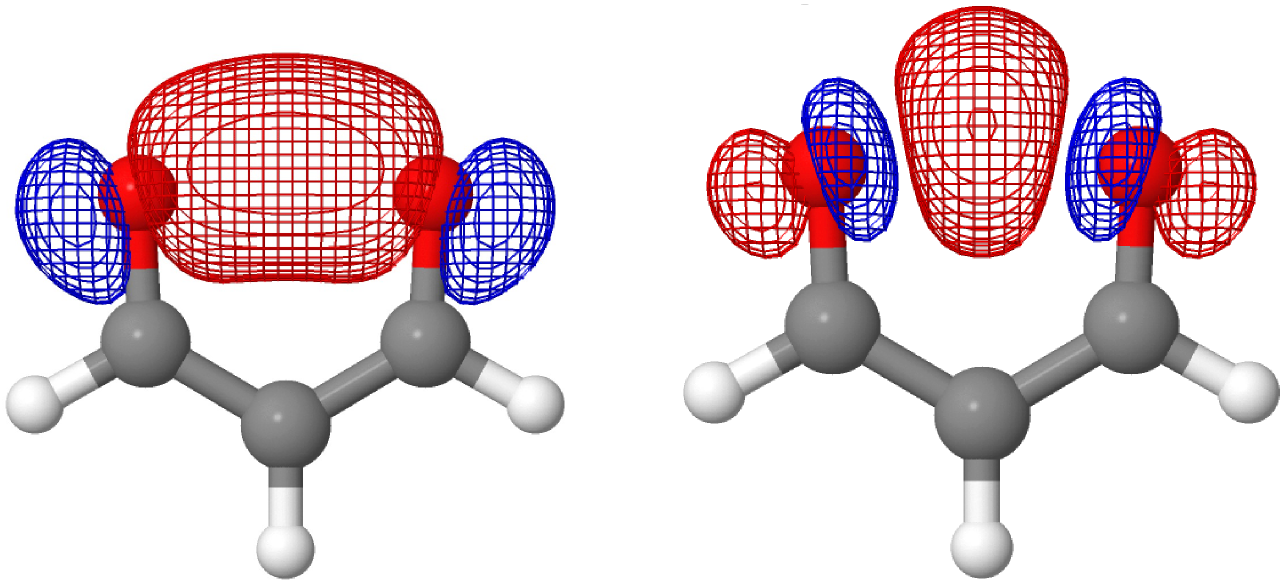}
    \caption{Electronic orbitals in the malonaldehyde active space.} 
    \label{fig:malon_act}
\end{figure}

\newpage

\section{\ce{OCHCO+} and \ce{FHF-} Protonic Densities}

\begin{figure}
    \centering
    \includegraphics[width=2.35in]{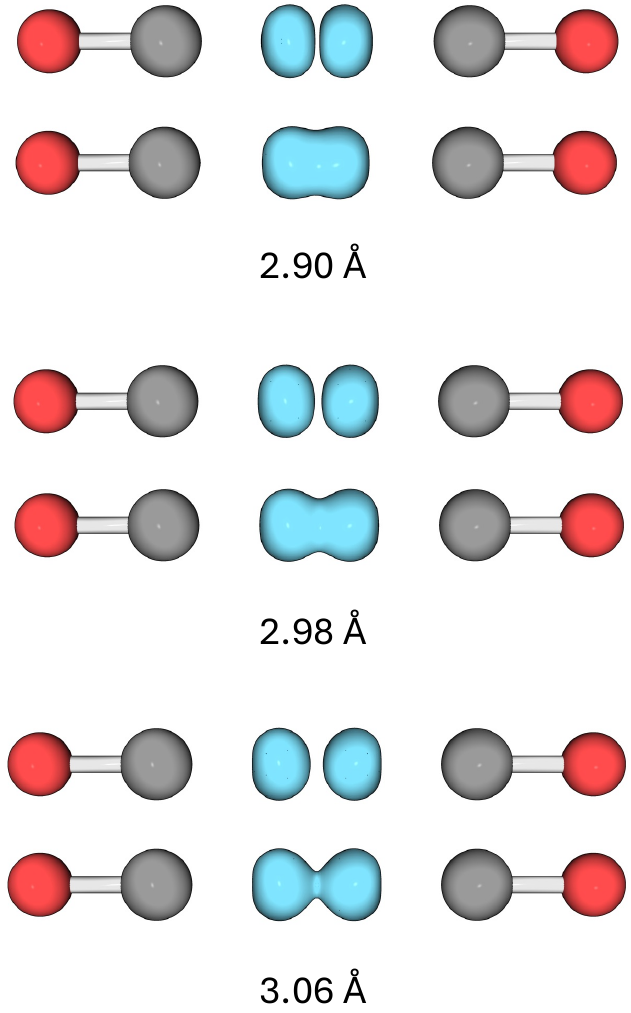}
    \caption{NEO-MR-SD$_\text{en}$CI protonic densities for \ce{OCHCO+} at the indicated C---C distances. The C atoms are shown in gray, the O atoms are shown in red, and the proton densities are shown in cyan with an isovalue of 0.04 e $\text{Bohr}^{-3}$.}
    \label{fig:ochco_densities}
\end{figure}

\begin{figure}
    \centering
    \includegraphics[width=6in]{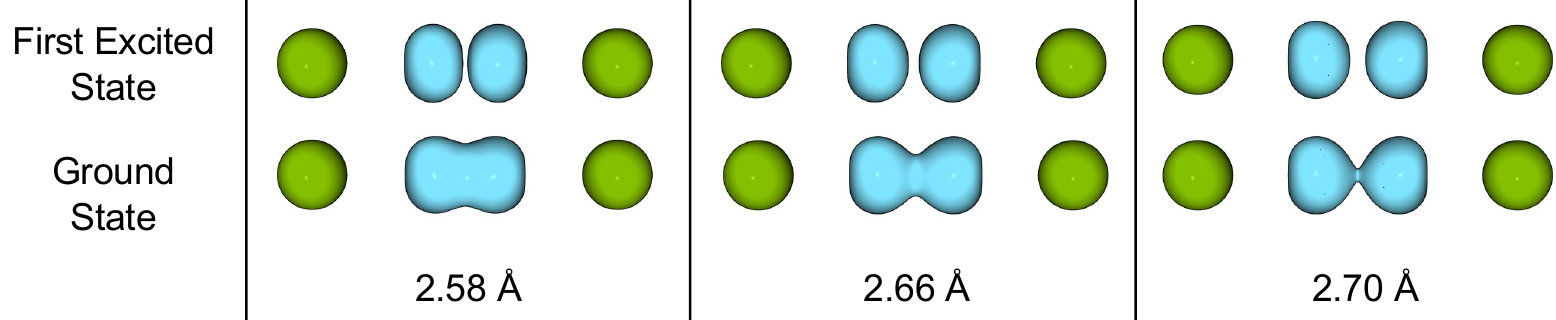}
    \caption{NEO-MR-SD$_\text{en}$CI protonic densities for \ce{FHF-} at the indicated F---F distances. The F atoms are shown in green, and the proton densities are shown in cyan with an isovalue of 0.04 e $\text{Bohr}^{-3}$.}
    \label{fig:fhf_densities}
\end{figure}

\section{Deuterium Basis Set Test}

\begin{table}[H]
\let\TPToverlap=\TPTrlap
\centering
\caption{\ce{FDF-} Tunneling Splitting (in \si{\wn}) Computed with NEO-MR-SD$_\text{en}$CI Using Unscaled and Scaled Protonic Basis Set \tnote{\emph{a}}}
\label{table:bas_set_conv_fhf}
\begin{threeparttable}
\begin{tabularx}{1\textwidth}{Y Y}
\toprule
\toprule
Protonic Basis Set & NEO-MR-SD$_\text{en}$CI\\
\midrule
8s8p8d8f & 70.63 \\
scaled 8s8p8d8f & 70.65 \\
\bottomrule
\end{tabularx}
\begin{tablenotes}\linespread{1}\footnotesize
    \item[\emph{a}] The F---F distance is 2.54 Å, and the exponents of the scaled protonic basis set are multiplied by $\sqrt{2}$.
\end{tablenotes}
\end{threeparttable}
\end{table}

\newpage
\section{Representative Timings of NEO-MR-SD$_\text{en}$CI Calculations}

Computational cost analysis for representative  calculations is provided in Tables \ref{table:mcscf_timings} and \ref{table:mrci_timings}. Note that the NEO-MCSCF and NEO-MRCI code has not been optimized, and these are not rigorous timing statistics. They are only intended to give readers a qualitative sense of the computational cost.

\begin{table}[H]
\let\TPToverlap=\TPTrlap
\centering
\caption{Timing Information for NEO-MCSCF \tnote{\emph{a}}}
\label{table:mcscf_timings}
\begin{threeparttable}
\begin{tabularx}{1\textwidth}{Y Y}
\toprule
\toprule
Molecule & CPU Time (hr) \\
\midrule
\ce{HeHHe+} & 2\\
\ce{FHF-} & 22 \\
\ce{OCHCO+} & 152 \\
Malonaldehyde & 144  \\
\bottomrule
\end{tabularx}
\begin{tablenotes}\linespread{1}\footnotesize
    \item[\emph{a}] Results obtained using the 8s8p8d8f protonic basis set for \ce{HeHHe+} and \ce{FHF-} and the 5s5p5d5f protonic basis set for \ce{OCHCO+} and malonaldehyde. The cc-pV5Z* electronic basis set was used on the protonic basis function centers. No electronic active space was used for \ce{HeHHe+}, the (2e,2o) electronic active space was used for \ce{FHF-}, \ce{OCHCO+}, and malonaldehyde, and a full protonic active space was used for all systems. The \ce{HeHHe+} geometry corresponds to a He---He distance of 2.40 Å; the \ce{FHF-} geometry corresponds to an F---F distance of 2.74 Å; the \ce{OCHCO+} geometry corresponds to a C---C distance of 3.06 Å; and the malonaldehyde geometry corresponds to an O---O distance of 2.55 Å.
\end{tablenotes}
\end{threeparttable}
\end{table}

\begin{table}[H]
\let\TPToverlap=\TPTrlap
\centering
\caption{Timing Information for NEO-MR-SD$_\text{en}$CI \tnote{\emph{a}}}
\label{table:mrci_timings}
\begin{threeparttable}
\begin{tabularx}{1\textwidth}{Y Y Y}
\toprule
\toprule
Molecule & CPU Time (hr) & NEO Configurations\\
\midrule
\ce{HeHHe+} & 4 & 125,868 \\
\ce{FHF-} & 186 & 3,670,776 \\
\ce{OCHCO+} & 334 & 5,990,720\\
Malonaldehyde & 445 & 8,314,240 \\
\bottomrule
\end{tabularx}
\begin{tablenotes}\linespread{1}\footnotesize
    \item[\emph{a}] Results obtained using the 8s8p8d8f protonic basis set for \ce{HeHHe+} and \ce{FHF-} and the 5s5p5d5f protonic basis set for \ce{OCHCO+} and malonaldehyde. The cc-pV5Z* electronic basis set was used on the protonic basis function centers. No electronic active space was used for \ce{HeHHe+}, the (2e,2o) electronic active space was used for \ce{FHF-}, \ce{OCHCO+}, and malonaldehyde, and a full protonic active space was used for all systems. The \ce{HeHHe+} geometry corresponds to a He---He distance of 2.40 Å; the \ce{FHF-} geometry corresponds to an F---F distance of 2.74 Å; the \ce{OCHCO+} geometry corresponds to a C---C distance of 3.06 Å; and the malonaldehyde geometry corresponds to an O---O distance of 2.55 Å.
\end{tablenotes}
\end{threeparttable}
\end{table}

\newpage
\section{Coordinates}

The geometries used in this work are provided below. These geometries include the optimized basis function center positions. The ghost centers are denoted with "GH". All units are in angstroms.

\subsection{\ce{HeHHe+}}

\noindent He---He distance 2.2 Å
\begin{table}[H]
\let\TPToverlap=\TPTrlap
    \centering
    \begin{tabularx}{1\textwidth}{>{\raggedleft\arraybackslash}X >{\raggedleft\arraybackslash}X >{\raggedleft\arraybackslash}X >{\raggedleft\arraybackslash}X}
        He & 0.00000000 & 0.00000000 & $-$1.10000000 \\
        He & 0.00000000 & 0.00000000 & 1.10000000 \\
        H & 0.00000000 & 0.00000000 &  $-$0.26938269 \\
        GH & 0.00000000 & 0.00000000 &  0.26938269 \\
    \end{tabularx}
\end{table}

\noindent He---He distance 2.25 Å
\begin{table}[H]
\let\TPToverlap=\TPTrlap
    \centering
    \begin{tabularx}{1\textwidth}{>{\raggedleft\arraybackslash}X >{\raggedleft\arraybackslash}X >{\raggedleft\arraybackslash}X >{\raggedleft\arraybackslash}X}
        He & 0.00000000 & 0.00000000 & $-$1.12500000 \\
        He & 0.00000000 & 0.00000000 & 1.12500000 \\
        H & 0.00000000 & 0.00000000 &  $-$0.30370531 \\
        GH & 0.00000000 & 0.00000000 &  0.30370531 \\
    \end{tabularx}
\end{table}

\noindent He---He distance 2.3 Å
\begin{table}[H]
\let\TPToverlap=\TPTrlap
    \centering
    \begin{tabularx}{1\textwidth}{>{\raggedleft\arraybackslash}X >{\raggedleft\arraybackslash}X >{\raggedleft\arraybackslash}X >{\raggedleft\arraybackslash}X}
        He & 0.00000000 & 0.00000000 & $-$1.15000000 \\
        He & 0.00000000 & 0.00000000 & 1.15000000 \\
        H & 0.00000000 & 0.00000000 &  $-$0.33611937 \\
        GH & 0.00000000 & 0.00000000 &  0.33611937 \\
    \end{tabularx}
\end{table}

\noindent He---He distance 2.35 Å
\begin{table}[H]
\let\TPToverlap=\TPTrlap
    \centering
    \begin{tabularx}{1\textwidth}{>{\raggedleft\arraybackslash}X >{\raggedleft\arraybackslash}X >{\raggedleft\arraybackslash}X >{\raggedleft\arraybackslash}X}
        He & 0.00000000 & 0.00000000 & $-$1.17500000 \\
        He & 0.00000000 & 0.00000000 & 1.17500000 \\
        H & 0.00000000 & 0.00000000 &  $-$0.36711178 \\
        GH & 0.00000000 & 0.00000000 &  0.36711178 \\
    \end{tabularx}
\end{table}

\noindent He---He distance 2.4 Å
\begin{table}[H]
\let\TPToverlap=\TPTrlap
    \centering
    \begin{tabularx}{1\textwidth}{>{\raggedleft\arraybackslash}X >{\raggedleft\arraybackslash}X >{\raggedleft\arraybackslash}X >{\raggedleft\arraybackslash}X}
        He & 0.00000000 & 0.00000000 & $-$1.12000000 \\
        He & 0.00000000 & 0.00000000 & 1.12000000 \\
        H & 0.00000000 & 0.00000000 &  $-$0.39618688 \\
        GH & 0.00000000 & 0.00000000 &  0.39618688 \\
    \end{tabularx}
\end{table}

\subsection{\ce{FHF-}}

\noindent F---F distance 2.54 Å
\begin{table}[H]
\let\TPToverlap=\TPTrlap
    \centering
    \begin{tabularx}{1\textwidth}{>{\raggedleft\arraybackslash}X >{\raggedleft\arraybackslash}X >{\raggedleft\arraybackslash}X >{\raggedleft\arraybackslash}X}
        F & 0.00000000 & 0.00000000 & $-$1.27000000 \\
        F & 0.00000000 & 0.00000000 & 1.27000000 \\
        H & 0.00000000 & 0.00000000 &  $-$0.25404100 \\
        GH & 0.00000000 & 0.00000000 &  0.25404100\\
    \end{tabularx}
\end{table}

\noindent F---F distance 2.58 Å
\begin{table}[H]
\let\TPToverlap=\TPTrlap
    \centering
    \begin{tabularx}{1\textwidth}{>{\raggedleft\arraybackslash}X >{\raggedleft\arraybackslash}X >{\raggedleft\arraybackslash}X >{\raggedleft\arraybackslash}X}
        F & 0.00000000 & 0.00000000 & $-$1.29000000 \\
        F & 0.00000000 & 0.00000000 & 1.29000000 \\
        H & 0.00000000 & 0.00000000 &  $-$0.28330200 \\
        GH & 0.00000000 & 0.00000000 &  0.28330200\\
    \end{tabularx}
\end{table}

\noindent F---F distance 2.62 Å
\begin{table}[H]
\let\TPToverlap=\TPTrlap
    \centering
    \begin{tabularx}{1\textwidth}{>{\raggedleft\arraybackslash}X >{\raggedleft\arraybackslash}X >{\raggedleft\arraybackslash}X >{\raggedleft\arraybackslash}X}
        F & 0.00000000 & 0.00000000 & $-$1.31000000 \\
        F & 0.00000000 & 0.00000000 & 1.31000000 \\
        H & 0.00000000 & 0.00000000 & $-$0.31109700 \\
        GH & 0.00000000 & 0.00000000 & 0.31109700 \\
    \end{tabularx}
\end{table}

\noindent F---F distance 2.66 Å
\begin{table}[H]
\let\TPToverlap=\TPTrlap
    \centering
    \begin{tabularx}{1\textwidth}{>{\raggedleft\arraybackslash}X >{\raggedleft\arraybackslash}X >{\raggedleft\arraybackslash}X >{\raggedleft\arraybackslash}X}
        F & 0.00000000 & 0.00000000 & $-$1.33000000 \\
        F & 0.00000000 & 0.00000000 & 1.33000000 \\
        H & 0.00000000 & 0.00000000 & $-$0.33774000 \\
        GH & 0.00000000 & 0.00000000 & 0.33774000 \\
    \end{tabularx}
\end{table}

\newpage

\noindent F---F distance 2.70 Å
\begin{table}[H]
\let\TPToverlap=\TPTrlap
    \centering
    \begin{tabularx}{1\textwidth}{>{\raggedleft\arraybackslash}X >{\raggedleft\arraybackslash}X >{\raggedleft\arraybackslash}X >{\raggedleft\arraybackslash}X}
        F & 0.00000000 & 0.00000000 & $-$1.35000000 \\
        F & 0.00000000 & 0.00000000 & 1.35000000 \\
        H & 0.00000000 & 0.00000000 & $-$0.36360540 \\
        GH & 0.00000000 & 0.00000000 & 0.36360540 \\
    \end{tabularx}
\end{table}

\noindent F---F distance 2.74 Å
\begin{table}[H]
\let\TPToverlap=\TPTrlap
    \centering
    \begin{tabularx}{1\textwidth}{>{\raggedleft\arraybackslash}X >{\raggedleft\arraybackslash}X >{\raggedleft\arraybackslash}X >{\raggedleft\arraybackslash}X}
        F & 0.00000000 & 0.00000000 & $-$1.37000000 \\
        F & 0.00000000 & 0.00000000 & 1.37000000 \\
        H & 0.00000000 & 0.00000000 & $-$0.38852230 \\
        GH & 0.00000000 & 0.00000000 & 0.38852230 \\
    \end{tabularx}
\end{table}

\subsection{\ce{OCHCO+}}

\noindent C---C distance 2.90 Å
\begin{table}[H]
\let\TPToverlap=\TPTrlap
    \centering
    \begin{tabularx}{1\textwidth}{>{\raggedleft\arraybackslash}X >{\raggedleft\arraybackslash}X >{\raggedleft\arraybackslash}X >{\raggedleft\arraybackslash}X}
        O & 0.00000000 & 0.00000000 & $-$2.56025300 \\
        O & 0.00000000 & 0.00000000 &  2.56025300 \\
        C & 0.00000000 & 0.00000000 &  $-$1.45000000 \\
        C & 0.00000000 & 0.00000000 &  1.45000000 \\
        H & 0.00000000 & 0.00000000 &  $-$0.28015112 \\
        GH & 0.00000000 & 0.00000000 &  0.28015112 \\
    \end{tabularx}
\end{table}

\noindent C---C distance 2.94 Å
\begin{table}[H]
\let\TPToverlap=\TPTrlap
    \centering
    \begin{tabularx}{1\textwidth}{>{\raggedleft\arraybackslash}X >{\raggedleft\arraybackslash}X >{\raggedleft\arraybackslash}X >{\raggedleft\arraybackslash}X}
        O & 0.00000000 & 0.00000000 & $-$2.58025300 \\
        O & 0.00000000 & 0.00000000 &  2.58025300 \\
        C & 0.00000000 & 0.00000000 &  $-$1.47000000 \\
        C & 0.00000000 & 0.00000000 &  1.47000000 \\
        H & 0.00000000 & 0.00000000 &  $-$0.30792752 \\
        GH & 0.00000000 & 0.00000000 &  0.30792752 \\
    \end{tabularx}
\end{table}

\newpage

\noindent C---C distance 2.98 Å
\begin{table}[H]
\let\TPToverlap=\TPTrlap
    \centering
    \begin{tabularx}{1\textwidth}{>{\raggedleft\arraybackslash}X >{\raggedleft\arraybackslash}X >{\raggedleft\arraybackslash}X >{\raggedleft\arraybackslash}X}
        O & 0.00000000 & 0.00000000 & $-$2.60025300 \\
        O & 0.00000000 & 0.00000000 &  2.60025300 \\
        C & 0.00000000 & 0.00000000 &  $-$1.49000000 \\
        C & 0.00000000 & 0.00000000 &  1.49000000 \\
        H & 0.00000000 & 0.00000000 &  $-$0.33383391 \\
        GH & 0.00000000 & 0.00000000 &  0.33383391 \\
    \end{tabularx}
\end{table}

\noindent C---C distance 3.02 Å
\begin{table}[H]
\let\TPToverlap=\TPTrlap
    \centering
    \begin{tabularx}{1\textwidth}{>{\raggedleft\arraybackslash}X >{\raggedleft\arraybackslash}X >{\raggedleft\arraybackslash}X >{\raggedleft\arraybackslash}X}
        O & 0.00000000 & 0.00000000 & $-$2.62025300 \\
        O & 0.00000000 & 0.00000000 &  2.62025300 \\
        C & 0.00000000 & 0.00000000 &  $-$1.51000000 \\
        C & 0.00000000 & 0.00000000 &  1.51000000 \\
        H & 0.00000000 & 0.00000000 &  $-$0.35990776 \\
        GH & 0.00000000 & 0.00000000 &  0.35990776 \\
    \end{tabularx}
\end{table}

\noindent C---C distance 3.06 Å
\begin{table}[H]
\let\TPToverlap=\TPTrlap
    \centering
    \begin{tabularx}{1\textwidth}{>{\raggedleft\arraybackslash}X >{\raggedleft\arraybackslash}X >{\raggedleft\arraybackslash}X >{\raggedleft\arraybackslash}X}
        O & 0.00000000 & 0.00000000 & $-$2.64025300 \\
        O & 0.00000000 & 0.00000000 &  2.64025300 \\
        C & 0.00000000 & 0.00000000 &  $-$1.53000000 \\
        C & 0.00000000 & 0.00000000 &  1.53000000 \\
        H & 0.00000000 & 0.00000000 &  $-$0.38528111 \\
        GH & 0.00000000 & 0.00000000 &  0.38528111 \\
    \end{tabularx}
\end{table}

\noindent C---C distance 3.10 Å
\begin{table}[H]
\let\TPToverlap=\TPTrlap
    \centering
    \begin{tabularx}{1\textwidth}{>{\raggedleft\arraybackslash}X >{\raggedleft\arraybackslash}X >{\raggedleft\arraybackslash}X >{\raggedleft\arraybackslash}X}
        O & 0.00000000 & 0.00000000 & $-$2.66025300 \\
        O & 0.00000000 & 0.00000000 &   2.66025300\\
        C & 0.00000000 & 0.00000000 &  $-$1.55000000 \\
        C & 0.00000000 & 0.00000000 &  1.55000000 \\
        H & 0.00000000 & 0.00000000 &  $-$0.40942460 \\
        GH & 0.00000000 & 0.00000000 &   0.40942460\\
    \end{tabularx}
\end{table}

\newpage

\subsection{Malondaldehyde}

\noindent O---O distance 2.55 Å
\begin{table}[H]
\let\TPToverlap=\TPTrlap
    \centering
    \begin{tabularx}{1\textwidth}{>{\raggedleft\arraybackslash}X >{\raggedleft\arraybackslash}X >{\raggedleft\arraybackslash}X >{\raggedleft\arraybackslash}X}
       GH  & $-$0.32439202 & $-$0.32293234 &  0.00000000 \\  
        H   &  0.32439202 & $-$0.32293234 &  0.00000000 \\  
        O   & $-$1.27500000 &  0.00000000 &  0.00000000 \\  
        O   &  1.27500000 &  0.00000000 &  0.00000000 \\  
        C   & $-$1.20470340 &  1.27297710 &  0.00000000 \\  
        C   &  1.20470340 &  1.27297710 &  0.00000000 \\  
        C   &  0.00000000 &  1.98626359 &  0.00000000 \\  
        H   &  0.00000000 &  3.06283598 &  0.00000000 \\  
        H   & $-$2.15371942 &  1.80885491 &  0.00000000 \\  
        H   &  2.15371942 &  1.80885491 &  0.00000000 \\  
    \end{tabularx}
\end{table}

\noindent O---O distance 2.62 Å
\begin{table}[H]
\let\TPToverlap=\TPTrlap
    \centering
    \begin{tabularx}{1\textwidth}{>{\raggedleft\arraybackslash}X >{\raggedleft\arraybackslash}X >{\raggedleft\arraybackslash}X >{\raggedleft\arraybackslash}X}
        GH  & $-$0.37316353 & $-$0.34058562 &  0.00000000 \\
        H   &  0.37316353 & $-$0.34058562 &  0.00000000 \\
        O   & $-$1.30832137 &  0.00507320 &  0.00000000 \\
        O   &  1.30832137 &  0.00507320 &  0.00000000 \\
        C   & $-$1.21275237 &  1.27683755 &  0.00000000 \\
        C   &  1.21275237 &  1.27683755 &  0.00000000 \\
        C   &  0.00000000 &  1.97816598 &  0.00000000 \\
        H   &  0.00000000 &  3.05507430 &  0.00000000 \\
        H   & $-$2.15184510 &  1.83064627 &  0.00000000 \\
        H   &  2.15184510 &  1.83064627 &  0.00000000 \\
    \end{tabularx}
\end{table}